\documentclass[aps,prx,twocolumn,secnumarabic,amsmath,amssymb,groupaddress]{revtex4-1}

\usepackage{titlesec}
\usepackage{bm}
\usepackage{comment}
\usepackage{verbatim}

\usepackage{graphicx}
\usepackage{subfigure}
\usepackage{tabularx}

\usepackage{color}
\usepackage[colorlinks,bookmarks=false,citecolor=darkblue,linkcolor=red,urlcolor=blue]{hyperref}
\def\cred{\color{red}}
\definecolor{darkred}{rgb}{0.7,0.0,0.0}
\def\cbl{\color{blue}}
\definecolor{darkblue}{rgb}{0,0.02,0.45}
\def\cdbl{\color{darkblue}}
\definecolor{darkgreen}{rgb}{0.02,0.45,0.0} 
\def\cgr{\color{darkgreen}}

\definecolor{violet}{rgb}{0.8,0.2,0.6}

\def\be{\begin{equation}}
\def\ee{\end{equation}}
\def\bea{\begin{eqnarray}}
\def\eea{\end{eqnarray}}

\def\vec{\mathbf}
\def\bs{\boldsymbol}
\def\mc{\mathcal}

\usepackage{times}

\begin{document}
\date{\today}

\title{Classical spin liquid instability driven by off-diagonal exchange in strong spin-orbit magnets}

\author{Ioannis Rousochatzakis}
\affiliation{School of Physics and Astronomy, University of Minnesota, Minneapolis, MN 55455, USA}

\author{N. B. Perkins}
\affiliation{School of Physics and Astronomy, University of Minnesota, Minneapolis, MN 55455, USA}

\begin{abstract}
We show that the off-diagonal exchange anisotropy drives Mott insulators with strong spin-orbit coupling to a classical spin liquid regime, characterized by an infinite number of ground states and Ising variables living on closed or open strings. Depending on the sign of the anisotropy, quantum fluctuations either fail to lift the degeneracy down to very low temperatures, or select non-collinear magnetic states with unconventional spin correlations. The results apply to all 2D and 3D tri-coordinated materials with bond-directional anisotropy, and provide a consistent interpretation of the suppression of the x-ray magnetic circular dichroism signal reported recently for $\beta$-Li$_2$IrO$_3$ under pressure.
\end{abstract}

\maketitle

\pagebreak

{\cdbl {\it Introduction --}}
The search for quantum spin liquids (QSLs) has been a central thread of correlated electron material research since their initial proposal several decades ago.~\cite{Anderson1973} Ideally, such systems evade magnetic order down to zero temperature and harbor a remarkable set of collective phenomena, including topological ground-state degeneracy, long-range entanglement, and fractionalized excitations.~\cite{HFMBook,Balents2010,Savary2016}
While the long activity on frustrated Mott insulators with 3$d$ ions has lead to a number of candidate QSLs with dominant isotropic interactions,~\cite{Balents2010} a certain class of 4$d$ and 5$d$ materials, the so-called Jackeli-Khaliullin Kitaev (JKK) systems,~\cite{Kitaev2006,Jackeli2009,Jackeli2010,Jackeli2013,Mandal2009,Obrien2016} with strong spin orbit coupling (SOC) and dominant anisotropic interactions has emerged in recent years as another prominent playground for QSLs.~\cite{Krempa2014} 
By now, several two- (2D) and three-dimensional (3D) materials have been identified in the JKK class, all with 3-fold coordinated magnetic ions that are well described by pseudo-spin $J_{\text{eff}}\!=\!1/2$ Kramer's doublets. Most notably, the layered A$_2$IrO$_3$ (A=Na,Li),~\cite{Singh2010,Singh2012,Liu2011,Choi2012,Ye2012,Chun2015,Williams2016} and $\alpha$-RuCl$_3$,~\cite{Plumb2014,Sears2015,Kubota2015,Majumder2015,Banerjee2016, Johnson2015} which are proximate to the honeycomb Kitaev QSL,~\cite{Kitaev2006} and the 3D harmonic-honeycomb Iridates $\beta$-Li$_2$IrO$_3$ and $\gamma$-Li$_2$IrO$_3$,~\cite{Biffin2014a,Biffin2014b,Modic2014,Takayama2015} which are proximate to generalized, exactly solvable Kitaev QSL's.~\cite{Mandal2009,Kimchi2014}

The key ingredients that lead to the desired degree of frustration in the JKK systems is the three-fold coordination of the magnetic sites and the nearest-neighbor (NN) Ising interactions along bond-dependent quantization axes. The compass form of this so-called Kitaev anisotropy stems from the highly entangled, spin-orbital nature of the Kramer's doublets.~\cite{Jackeli2010,Katukuri2014,Sizyuk2014,Rau2014,Shankar2014,Katerina2013,Yamaji2014,Kim2015,Winter2016,Kee2016,Kim2016} 
While this anisotropy seems to be the dominant interaction in all JKK materials, experiments show that these systems order magnetically at sufficiently low temperatures,~\cite{Singh2010,Singh2012,Liu2011,Choi2012,Ye2012,Chun2015,Williams2016,Biffin2014a,Biffin2014b,Modic2014,Takayama2015,Plumb2014,Sears2015,Kubota2015,Majumder2015,Banerjee2016, Johnson2015} consistent with theoretical predictions that Kitaev QSLs are fragile against weak perturbations.~\cite{Jackeli2010,Jackeli2013,Schaffer2012,Lee2014,Katukuri2014,Satoshi2016,Katukuri2015,Ioannis2015} 

Nevertheless, the aspiration to find spin liquid physics in the JKK systems still stands. The new experimental direction is to study these materials under external perturbations, such as magnetic field,~\cite{Buchner2016} chemical substitution,~\cite{Takagi2016} and hydrostatic pressure.~\cite{Takayama2015} For $\beta$-Li$_2$IrO$_3$, for example, x-ray magnetic circular dichroism (XMCD) data show a strong reduction of the magnetic-field-induced ferromagnetic moments with pressure, and a complete suppression around 2 GPa.~\cite{Takayama2015} Since the system remains insulating under pressure, the authors suggest that the vanishing of the XMCD signal reflects that the system is driven into a spin-liquid regime. 

The natural interpretation of these results would be that pressure brings $\beta$-Li$_2$IrO$_3$ closer to the Kitaev QSL. Surprisingly, however, there are two independent {\it ab initio} studies, one from density functional theory~\cite{Kim2016} and another from quantum chemistry,~\cite{Liviu2016} showing that, under pressure, the system actually departs further from the ideal Kitaev model, and that the interaction that becomes increasingly relevant is the symmetric off-diagonal exchange $\Gamma$.~\cite{Katukuri2014,Rau2014,Satoshi2016,Sizyuk2014,Winter2016,Kee2016}

Motivated by the above reports, we set out to investigate the physics of the JKK systems in the region where $\Gamma$ is the dominant coupling. 
Remarkably, the key qualitative results are shared by all 2D and 3D JKK systems. The $\Gamma$ coupling drives these systems toward a classical spin liquid regime, characterized by an infinite number of classical ground states. This degeneracy is not accidental but arises from an infinite number of zero- and/or one-dimensional gauge symmetries that exist only for classical spins. For quantum spins, the degeneracy is eventually lifted by the order-by-disorder mechanism at an energy scale which depends strongly on the sign of $\Gamma$. For $\Gamma\!>\!0$, the leading quantum fluctuations fail to remove the frustration, giving rise to a `cooperative paramagnet' down to very low temperatures. For $\Gamma\!<\!0$, fluctuations select a multi-sublattice, non-collinear state with vanishing total moment. Both scenarios are consistent with the suppression of the XMCD signal under pressure, although the latter might be more relevant for $\beta$-Li$_2$IrO$_3$, according to {\it ab initio} studies.~\cite{Kim2016,Liviu2016}

{\cdbl {\it Model --}}
The JKK systems have three different types of NN bonds, labeled as $\alpha\!=\!x$, $y$, or $z$, shown schematically as  
\be\label{gr:3bonds}
\includegraphics[width=0.7in]{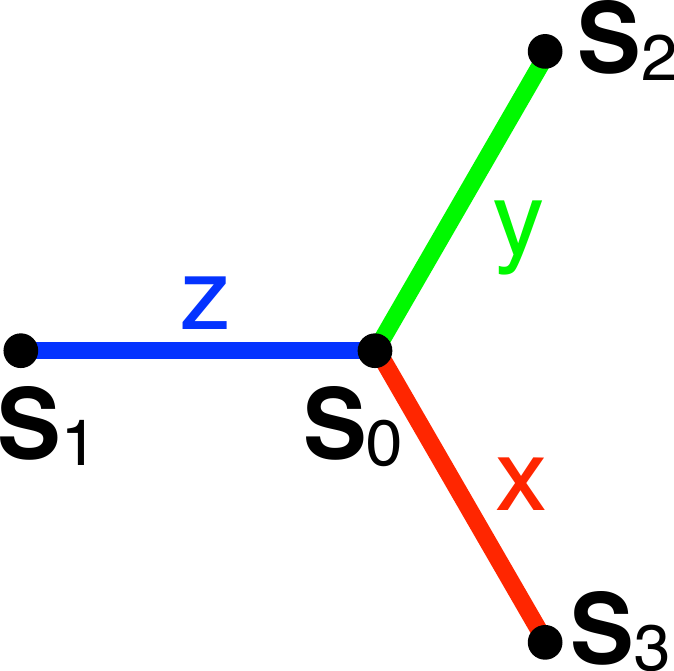}
\ee 
where $\vec{S}_i$ denotes the pseudospin 1/2 residing at the vertex $i$. 
The Hamiltonian describing the symmetric part of the off-diagonal exchange anisotropy reads
\bea\label{eq:HGamma1}
\mc{H}\!&=&\!\!\Gamma \!\!\!\sum_{\langle ij\rangle \in \text{`x'}}\!\!\!(S_i^yS_j^z+S_i^zS_j^x)
\pm \Gamma\!\!\! \sum_{\langle ij\rangle \in \text{`y'}}\!\!\!(S_i^zS_j^x+S_i^xS_j^z) \nonumber\\
&&\hspace{0.5cm}\pm\Gamma\!\!\! \sum_{\langle ij\rangle \in \text{`z'}}\!\!\!(S_i^xS_j^y+S_i^yS_j^x),
\eea
where $\langle ij\rangle$ denotes NN sites, and $\pm$ accounts for the sign modulation of the couplings on $x$- and $y$-bonds in the 3D systems.~\cite{Lee2016} For the 2D case all bonds have the plus sign.

\begin{figure}[!t] 
\includegraphics[width=0.4\textwidth,angle=0,clip=true,trim=0 0 0 0]{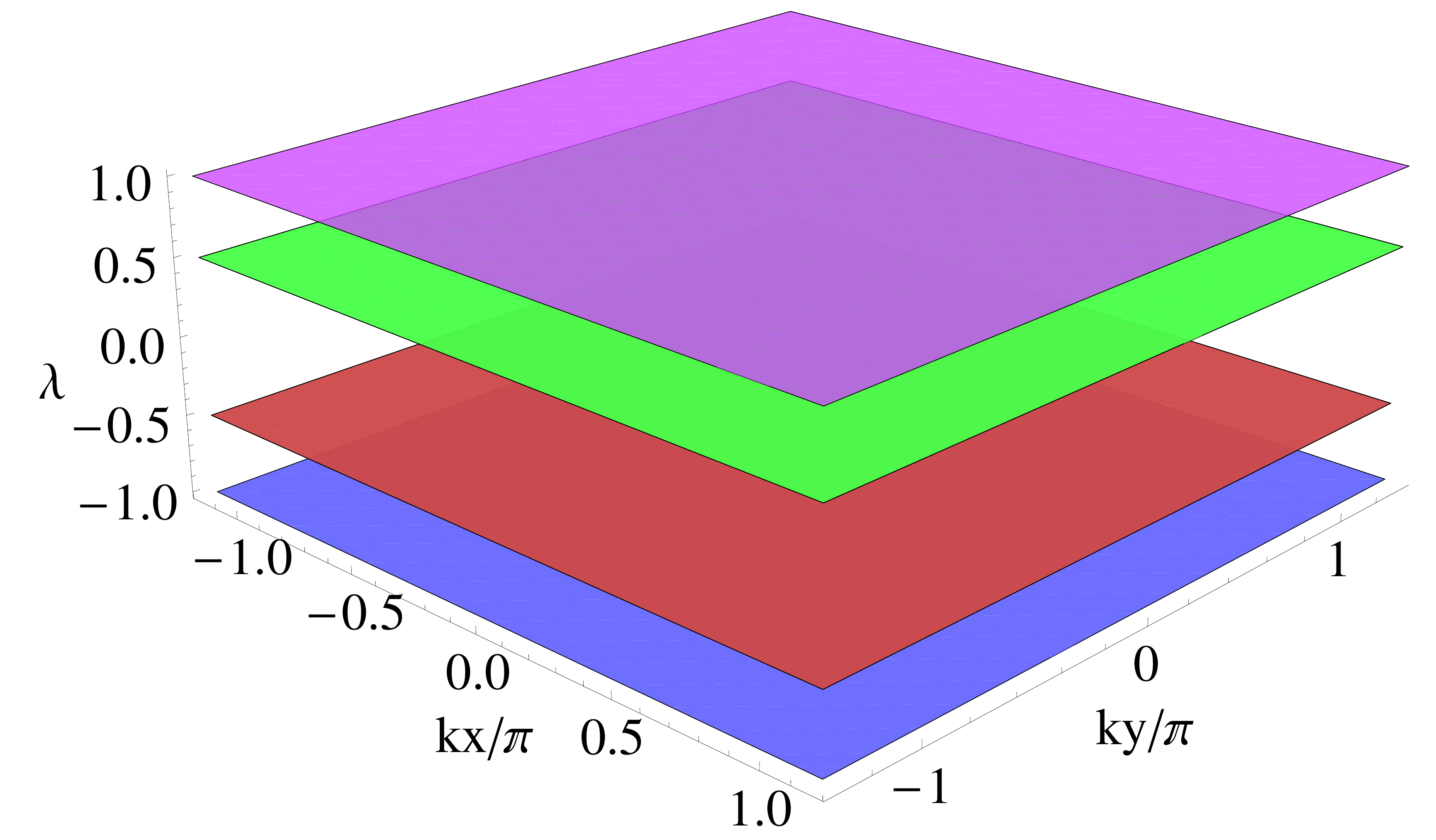}
\caption{Spectrum $\lambda_{1-6}/|\Gamma|$ of the matrix $\bs{\Lambda}_{\vec{k}}$ entering the Fourier transform of the classical energy, see Supplementary material.~\cite{SM}}\label{fig:flatbands}
\end{figure}

{\cdbl {\it Classical limit --}}
Let us consider the classical limit where $\vec{S}_i$ are vectors of length $S$, and begin with the simplest 2D honeycomb case (we generalize to 3D below). 
The highly frustrated nature of this model is first revealed by the fact that the lowest eigenvalue of the 6$\times$6 interaction matrix $\bs{\Lambda}_\vec{k}$ in momentum space~\cite{SM} is completely flat. In fact, the same is true for all six bands, see Fig.\ref{fig:flatbands}. Specifically, $\lambda_1\!=\!-|\Gamma|$, $\lambda_2\!=\!\lambda_3\!=\!-|\Gamma|/2$, $\lambda_4\!=\!\lambda_5\!=\!|\Gamma|/2$, and $\lambda_6\!=\!|\Gamma|$.

To understand the structure of the ground states and why there is an infinite number of them, we search for states that saturate the lower bound of the energy per site $\lambda_1 S^2$.~\cite{SM} 
Consider a pair of NN spins, say $\vec{S}_0$ and $\vec{S}_1$ of (\ref{gr:3bonds}), which interact with a term $\Gamma \left( S_0^xS_1^y+S_0^yS_1^x\right)$. If these spins were isolated from the rest, then their mutual energy would be minimized by placing the spins on the $xy$-plane with $S_1^x\!=\!\zeta S_0^y$, $S_1^y\!=\!\zeta S_0^x$, where $\zeta\!=\!-\text{sgn}(\Gamma)$. Similarly, for the $x$-bond of (\ref{gr:3bonds}), we would get $S_3^y\!=\!\zeta S_0^z$, $S_3^z\!=\!\zeta S_0^y$, and for the $y$-bond of (\ref{gr:3bonds}), $S_2^x\!=\!\zeta S_0^z$, $S_2^z\!=\!\zeta S_0^x$.  Returning to the lattice problem, the idea is to require that the two components involved in each $\Gamma$ term satisfy the respective relations above, without specifying the third component for the moment. This is done as follows: (i) We choose a direction for the central spin of (\ref{gr:3bonds}) and parametrize it as
\be
\vec{S}_0=(\eta_1 {\cred a}, \eta_2 {\cgr b}, \eta_3 {\cbl c}),
\ee
where ${\cred a}\!=\! |S_0^x|$, ${\cgr b}\!=\!|S_0^y|$, ${\cbl c}\!=\!|S_0^z|$,  $\eta_1\!=\!\text{sgn}(S_0^x)$, $\eta_2\!=\!\text{sgn}(S_0^y)$ and $\eta_3\!=\!\text{sgn}(S_0^z)$. 
Then, (ii) we fix two components of the three neighboring spins as follows:
\be
\begin{array}{c}
\vec{S}_1=(\zeta \eta_2 {\cgr b},\zeta \eta_1 {\cred a},S_1^z),~~~
\vec{S}_2=(\zeta \eta_3 {\cbl c},S_2^y,\zeta \eta_1 {\cred a}),\\
\vec{S}_3=(S_3^x,\zeta \eta_3 {\cbl c},\zeta \eta_2 {\cgr b}).
\end{array}
\ee
Then, (iii) we fix accordingly two components of the neighbors of $\vec{S}_1$, $\vec{S}_2$, and $\vec{S}_3$, and so on, until we cover the whole lattice. It is easy to see that the total energy of the generated configurations saturates the lower energy bound, and are therefore ground states. Indeed, the contribution to the energy from the cluster (\ref{gr:3bonds}) is
$E_{\includegraphics[width=0.1in]{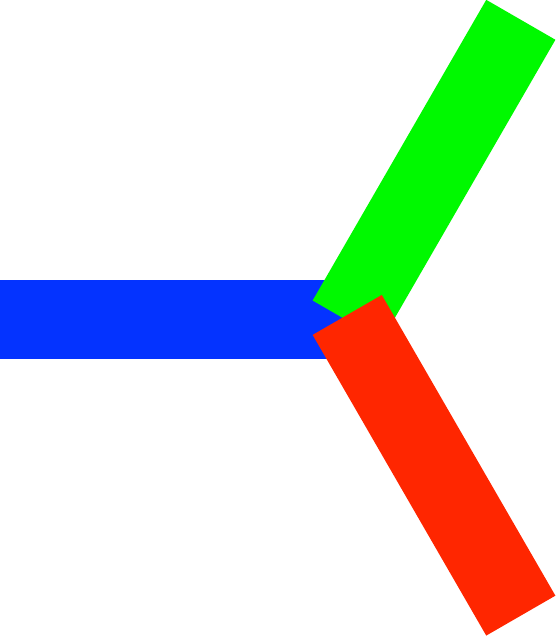}} \!=\! -2 ({\cred a}^2+{\cgr b}^2+{\cbl c}^2) |\Gamma| \!=\! -2|\Gamma| S^2$, 
and the same is true for any such cluster in the lattice. Since each bond is shared by two sites, the total energy per site is $E/N\!=\!-|\Gamma| S^2$, which saturates the lower energy bound.

\begin{figure}[!t] 
\includegraphics[width=0.36\textwidth,angle=0,clip=true,trim=0 0 0 0]{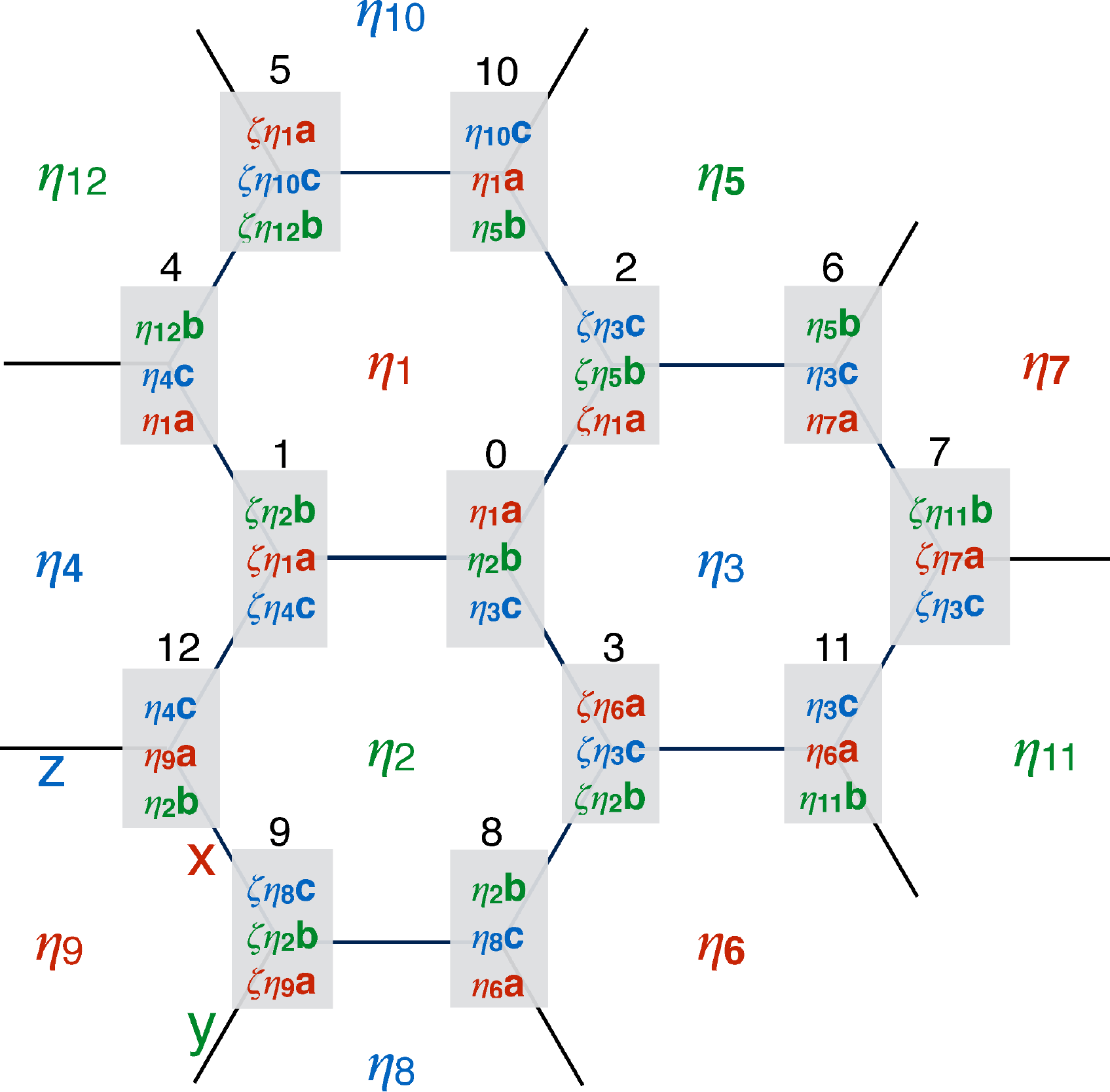}
\caption{Classical ground states of the $\Gamma$ model on the 2D honeycomb lattice, where ${\cred a}^2+{\cgr b}^2+{\cbl c}^2\!=\!S^2$ and $\eta_i\!=\!\pm1$.}\label{fig:GS}
\end{figure}

\begin{figure*}[!t] 
\includegraphics[width=0.4\textwidth,angle=0,clip=true,trim=0 0 0 0]{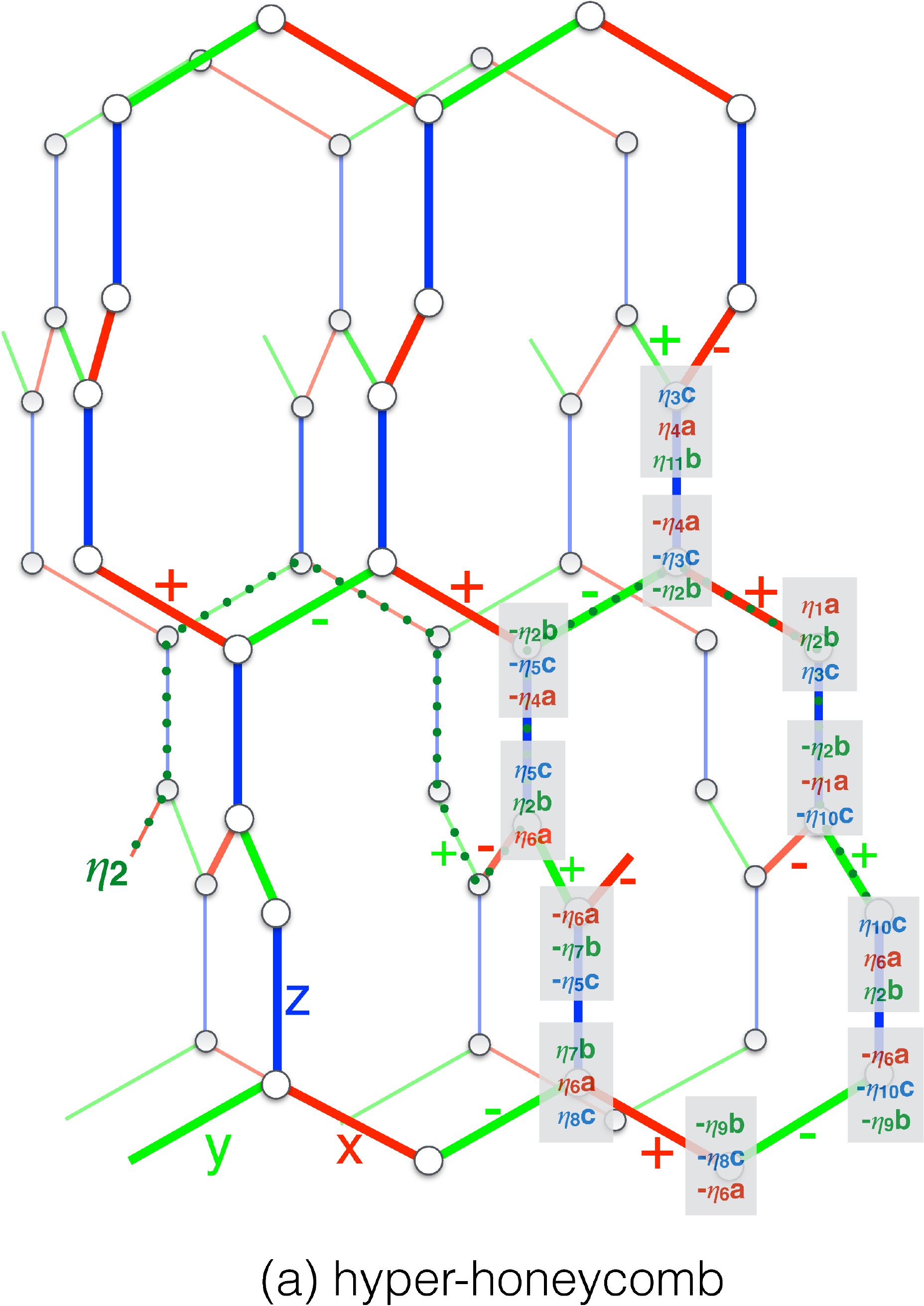}~~~~~~
\includegraphics[width=0.43\textwidth,angle=0,clip=true,trim=0 0 0 0]{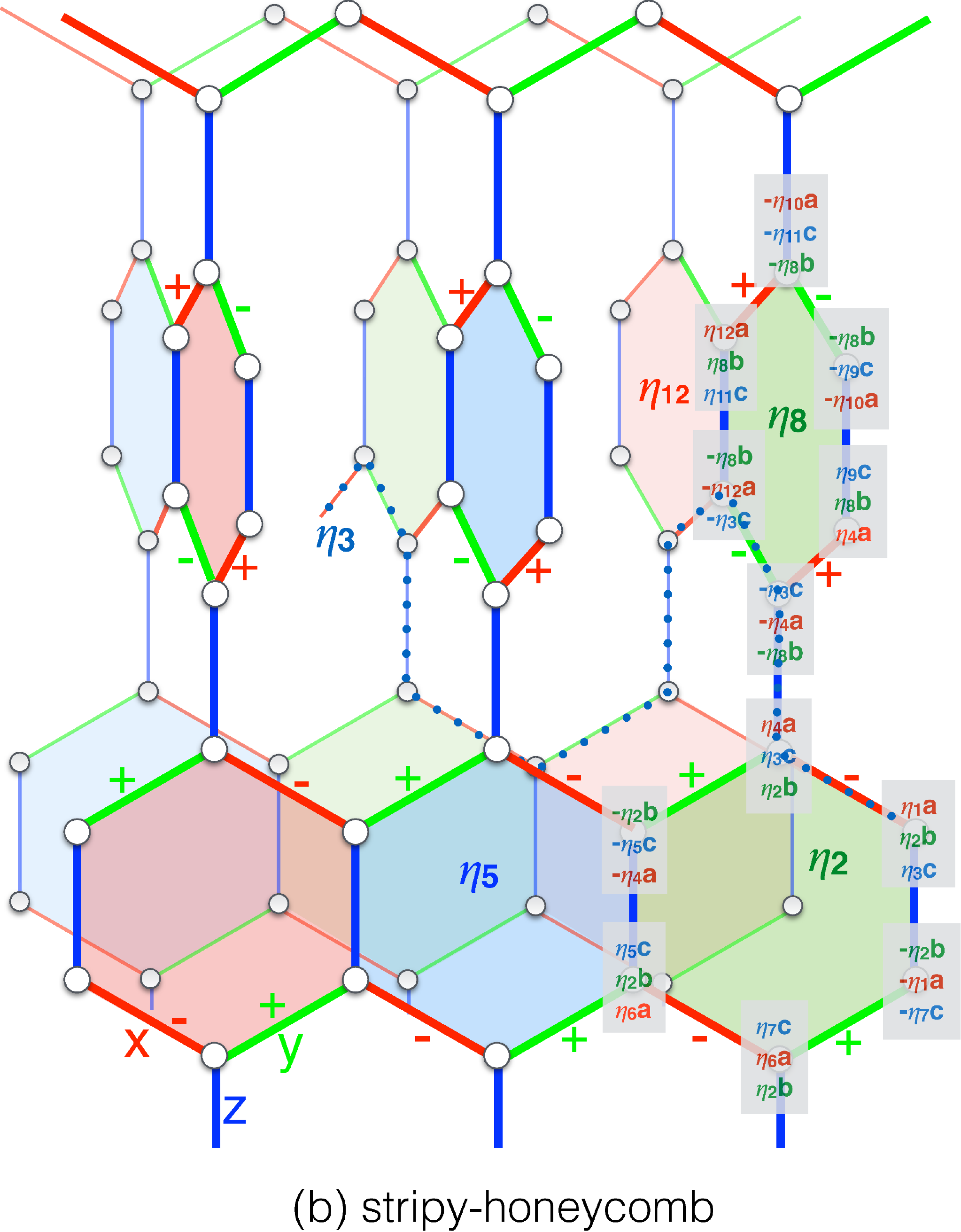}
\caption{Classical ground states of the $\Gamma$ model on $\beta$-Li$_2$IrO$_3$ (a) and $\gamma$-Li$_2$IrO$_3$ (b), for $\Gamma\!>\!0$. The $\pm$ signs labeling the $x$ or $y$ bonds denote the signs of the associated $\Gamma$ coupling relative to that on the $z$ bonds.~\cite{Lee2016} The dotted strings show the open strings where $\eta_2$ (a) and $\eta_3$ (b) live.
}\label{fig:3D}
\end{figure*}

Now, the reason why there are infinite ground states lies in the freedom to choose the third component of the spins, i.e., $S_1^z$, $S_2^y$, $S_3^x$, etc. Imposing the spin length constraint shows that this freedom is associated with the signs of these components: 
\be
S_1^z=\zeta \eta_4 {\cbl c}, ~S_2^y=\zeta \eta_5 {\cgr b}, ~S_3^x=\zeta \eta_6 {\cred a}, 
\ee
where $\eta_i\!=\!\pm 1$ are Ising-like variables. The choice of signs in front of the $\eta$'s give the simplest representation of the state as we see below, but is otherwise arbitrary. To find out how many independent $\eta$'s exist, we examine more closely what happens around the central cluster (\ref{gr:3bonds}), see Fig.~\ref{fig:GS}. This picture shows that each $\eta_i$ appears only around a single hexagon, and so we can label the ground states by assigning the $\eta$'s to the hexagons. This parametrization in terms of {\it local} Ising variables gives a total of $2^{N/2}$ states for a fixed choice of $\{{\cred a},{\cgr b},{\cbl c}\}$. Note that if two (or one) of $\{{\cred a},{\cgr b},{\cbl c}\}$ vanish then 2/3 (resp. 1/3) of the $\eta$'s are idle and we get $2^{N/6}$ (resp. $2^{N/3}$) states instead. On top of this degeneracy, there is also the continuous degeneracy associated to the choice of $\{{\cred a}, {\cgr b}, {\cbl c}\}$. 

The $\eta$-parametrization reveals that the local zero-energy modes responsible for the extensive degeneracy correspond to flipping one particular component of each of the six spins of a hexagon. For the $\eta_1$ hexagon of Fig.~(\ref{fig:GS}), for example, the zero mode amounts to simultaneously flipping the signs of $S_0^x$, $S_1^y$, $S_4^z$, $S_5^x$, $S_{10}^y$, and $S_2^z$. This operation is in fact a symmetry of the classical Hamiltonian, so the ground state degeneracy associated with the $\eta$'s is not accidental but symmetry related. Inspecting the form of the $\Gamma$ terms, these symmetries involve strings of alternating $x$-$y$-$z$ bonds which happen to be hexagons in the 2D honeycomb case. We shall come back to this when we discuss the 3D cases below. 

Another key aspect of the $\eta$ variables is that they split into three inequivalent types that occupy the vertices of three inter-penetrating triangular sublattices ${\cred A}$, ${\cgr B}$ and $ {\cbl C}$ (denoted by red, green and blue in Fig.~\ref{fig:GS}). Type-${\cred A}$ (resp. ${\cgr B}$, $ {\cbl C}$) hexagons are characterized by alternating spin components with magnitude ${\cred a}$ (resp. ${\cgr b}$, ${\cbl c})$. This structure is reflected directly in the values of the so-called fluxes $\{W_h\}$, that are known from the quantum Kitaev model.~\cite{Kitaev2006} Indeed, from Fig.~\ref{fig:GS}: 
\bea
&&
W_{h\in {\cred A}}=W_{\eta_1}\!\!=\! S_0^x S_1^y S_4^z S_5^x S_{10}^y S_2^z/S^6 \!=\! \zeta{\cred \tilde{a}}^6,~~\\
&&
W_{h\in {\cgr B}}=W_{\eta_2}\!\!=\! S_8^x S_9^y S_{12}^z S_{1}^x S_{0}^y S_3^z/S^6\!=\! \zeta{\cgr \tilde{b}}^6,~~ \\
&&
W_{h\in {\cbl C}}=W_{\eta_3}\!\!=\! S_{11}^x S_3^y S_0^z S_2^x S_6^y S_7^z/S^6 \!=\! \zeta{\cbl \tilde{c}}^6,  
\eea
where ${\cred\tilde{a}}\!=\!{\cred a}/S$, ${\cgr\tilde{b}}\!=\!{\cgr b}/S$ and ${\cbl\tilde{c}}\!=\!{\cbl c}/S$. This flux pattern is shared by all ground states with fixed $\{{\cred a}, {\cgr b}, {\cbl c}\}$.~\footnote{Note that here, in contrast to the Kitaev model, the fluxes are not conserved quantities, and so they cannot be fixed independently from each other.} 
The most striking manifestation of the three-sublattice structure of the $\eta$'s, however, shows up when we take into account quantum fluctuations, see below.

The above steps can be repeated for both $\beta$-Li$_2$IrO$_3$ and $\gamma$-Li$_2$IrO$_3$, see Fig.~\ref{fig:3D}. There are again infinite ground states characterized by Ising variables $\eta$ of three types, as in 2D. There is however one qualitative difference in the nature of the zero-energy modes which stems from the way alternating $x$-$y$-$z$ bonds propagate in the lattice. In $\beta$-Li$_2$IrO$_3$, they form infinite strings, so all $\eta$'s are {\it nonlocal} [see e.g. the ${\cgr\eta_2}$ string in Fig.~\ref{fig:3D}~(a)] and the degeneracy is sub-extensive. In $\gamma$-Li$_2$IrO$_3$, the alternating $x$-$y$-$z$ bonds form either closed hexagons or infinite strings. Hence, some $\eta$'s are local (giving an extensive degeneracy), like ${\cred\eta_1}$, ${\cred\eta_{12}}$, ${\cgr\eta_2}$, ${\cgr\eta_8}$, ${\cbl\eta_5}$ and ${\cbl\eta_9}$ in Fig.~\ref{fig:3D}~(b), but the rest live on open strings, like ${\cbl\eta_3}$. So, $\gamma$-Li$_2$IrO$_3$ is intermediate between the 2D honeycomb and $\beta$-Li$_2$IrO$_3$.

{\cdbl {\it Quantum order-by-disorder --}}
The above zero- and one-dimensional gauge symmetries that are responsible for the zero-energy modes are very common in compass-like models and act to suppress local magnetic order by virtue of a generalized Elitzur's theorem.~\cite{Batista05,Nussinov2006,Nussinov2015} Here, however, these symmetries exist only for classical spins, because they involve time reversal and affect only part of the system (a hexagon or an open string). For quantum spins such operations cannot be effected (because time reversal is global), meaning that the classical degeneracy is lifted and local order is possible. This leads us to the important question of order-by-disorder, which we address here by real space perturbation theory (RSPT).~\cite{Lindgard1988,Long1989,Heinila1993,Mike2014} In this approach, one introduces local axes $\vec{e}_i^z$ along the classical spin directions, and then splits $\mc{H}$ into a diagonal part $\mc{H}_0\!=\!h\sum_i (S\!-\!\vec{S}_i\cdot\vec{e}_i^z)$, describing fluctuations in the local field $h\!=\!2|\Gamma|S$, and a perturbation $\mc{V}\!=\!\mc{H}\!-\!\mc{H}_0$, which couples fluctuations on different sites.~\cite{SM}

It turns out that the order-by-disorder physics can be captured already by the leading, short-wavelength spin-wave corrections, which can be obtained from second-order perturbation theory on isolated bonds. The corrections from the three types of bonds, say ($\vec{S}_0$,$\vec{S}_3$), ($\vec{S}_0$,$\vec{S}_2$) and ($\vec{S}_0$,$\vec{S}_1$) of Fig.~(\ref{fig:GS}), are (disregarding constants):
\be \label{eq:2order}
\begin{array}{c}
\delta E_{03}=(\Gamma S {\cred \tilde{a}}^2/8)  {\cred \eta_1\eta_6} -|\Gamma| S{\cred \tilde{a}}^4/16,\\
\delta E_{02}=(\Gamma S {\cgr \tilde{b}}^2/8)  {\cgr\eta_2\eta_5} -|\Gamma| S {\cgr \tilde{b}}^4/16,\\
\delta E_{01}=(\Gamma S {\cbl \tilde{c}}^2/8)  {\cbl\eta_3\eta_4} -|\Gamma| S {\cbl \tilde{c}}^4/16 .
\end{array}
\ee
These expressions give two important insights:

(i) The correction from each bond type $\alpha$ does not depend on all four $\eta$'s involved in the bond, but only on the ones involved in the $\alpha$-th component of the spins. As a result, different types of $\eta$'s do not couple to each other. This is a consequence of three gauge-like global symmetries which amount to flipping the sign of all $\eta$'s of a given type.~\cite{SM} A coupling between different type of $\eta$'s eventually arises in fourth order (from connected, three-site or larger clusters), but this coupling is much smaller, see below.

Within each $\eta$-sublattice then, we obtain an effective Ising model with coupling $J_{{\cred A}}\!=\!\Gamma S {\cred \tilde{a}}^2/8$, $J_{{\cgr B}}\!=\!\Gamma S {\cgr \tilde{b}}^2/8$, or $J_{{\cbl C}}\!=\!\Gamma S {\cbl \tilde{c}}^2/8$.
Remarkably, when $\Gamma\!>\!0$, these models are highly frustrated for all 2D and 3D cases. 
For the 2D honeycomb, each $\eta$-sublattice is described by a triangular Ising antiferromagnet, the prototype of classical spin liquids.~\cite{Wannier1950,*Houtappel1950} 
For the hyper-honeycomb, the frustration arises again from AF Ising `triangles' of $\eta$ variables, such as $\{{\cred \eta_1}, {\cred\eta_4}, {\cred\eta_6}\}$ or $\{{\cgr\eta_2}, {\cgr\eta_7},  {\cgr\eta_9}\}$ or $\{{\cbl\eta_5}, {\cbl\eta_8}, {\cbl\eta_{10}}\}$ in Fig.~\ref{fig:3D}~(a).
Such effective triangles occur at the length-ten loops of the lattice, where the corresponding open strings pass nearby each other. 
The same is true for the nonlocal $\eta$'s in the stripy-honeycomb, where two effective triangles such as $\{{\cred \eta_1}, {\cred\eta_4}, {\cred\eta_6}\}$ and $\{{\cbl\eta_3}, {\cbl\eta_5}, {\cbl\eta_7}\}$ in Fig.~\ref{fig:3D}~(b), are formed at a hexagon (${\cgr \eta_2}$) of the complementary color. At the same time, the local variables form 1D AF chains (formed by hexagons), and there is also a frustrating coupling between local and nonlocal $\eta$'s. 
So, in all 2D and 3D $\Gamma$ models, there is strong frustration within each $\eta$-sublattice when $\Gamma\!>\!0$. 

(ii) Whatever the ground state within each $\eta$-sublattice is, the dependence of the total energy on $\{{\cred a}, {\cgr b}, {\cbl c}\}$ through $J_{{\cred A}}$, $J_{{\cgr B}}$ and $J_{{\cbl C}}$  drops out because the three sublattices have identical $\langle \eta \eta'\rangle$ correlations and because ${\cred a}^2\!+\!{\cgr b}^2\!+\!{\cbl c}^2\!=\!S^2$. However, the second terms of Eq.~(\ref{eq:2order}) give rise to a fourth-order cubic anisotropy $E_{\text{ani}}$, of the form 
\be\label{eq:Eani}
E_{\text{ani}}/N=-\frac{|\Gamma| S}{32}({\cred\tilde{a}}^4+{\cgr\tilde{b}}^4+{\cbl\tilde{c}}^4),
\ee
which is the leading mechanism by which the system lifts the degeneracy associated to the choice of $\{{\cred a}, {\cgr b}, {\cbl c}\}$. Here,  $E_{\text{ani}}$ is minimized along the cubic axes, i.e. when $\{{\cred\tilde{a}}, {\cgr\tilde{b}}, {\cbl\tilde{c}}\}\!=\!\{1,0,0\}$, $\{0,1,0\}$ or $\{0,0,1\}$. 
When this happens, 2/3 of the $\eta$'s become idle and only the behavior of the remaining 1/3 variables has to be understood. 

For $\Gamma\!>\!0$, the systems remain highly frustrated even well below the energy scale set by $E_{\text{ani}}$. Residual corrections eventually stabilize some type of order but only at a much smaller energy scale. For the 2D honeycomb case, for example, tunneling processes between different classical ground states give rise to transverse corrections to the  Ising Hamiltonian, leading to an XYZ model and a peculiar state with two order parameters, one magnetic and one nematic.~\cite{IoannisLong} 

For $\Gamma\!<\!0$, the systems order magnetically below an energy scale set by $E_{\text{ani}}$. The order corresponds to a FM alignment of the $\eta$ variables of one type (the other two become idle). In terms of the underlying spins, this state has a multi-sublattice non-collinear structure, with spins pointing along the cubic axes. The 2D honeycomb has three spin sublattices and a finite total moment along $[111]$. The 3D systems, on the other hand, have six spin sublattices due to the modulation of the relative signs of $\Gamma$ on $x$- and $y$-bonds, see Fig.~\ref{fig:3D}. So, the ordered state of the 3D systems for $\Gamma\!<\!0$ has zero total moment.

{\cdbl {\it Higher-order terms --}} To highlight the unimportance of higher-order corrections we report here the fourth-order RSPT corrections on the connected cluster (\ref{gr:3bonds}), for the most quantum case of $S\!=\!1/2$ (expressions for $S\!>\!\frac{1}{2}$ are given in \cite{SM}). The corrections to the bilinear couplings are
$\delta J_{{\cred A}}\!=\!\frac{\Gamma}{768} (-14{\cred\tilde{a}}^2\!+\!33{\cred\tilde{a}}^4\!-\!25{\cred\tilde{a}}^6\!+\!23{\cred\tilde{a}}^2{\cgr\tilde{b}}^2{\cbl\tilde{c}}^2)$, and similarly for $\delta J_{{\cgr B}}$ and $\delta J_{{\cbl C}}$ by cyclic permuting $\{{\cred\tilde{a}},{\cgr\tilde{b}},{\cbl\tilde{c}}\}$. These corrections are very small (at maximum they are only $1/8$ of the second-order couplings).
Next, the corrections to $E_{\text{ani}}$ are $\delta E_{\text{ani}}\!=\!\frac{-|\Gamma|}{3072} [15({\cred\tilde{a}}^4\!+\!{\cgr\tilde{b}}^4\!+\!{\cbl\tilde{c}}^4)\!+\!95{\cred\tilde{a}}^2{\cgr\tilde{b}}^2{\cbl\tilde{c}}^2 \!-\!22({\cred\tilde{a}}^8\!+\!{\cgr\tilde{b}}^8\!+\!{\cbl\tilde{c}}^8)]$. The 6th- and 8th-order anisotropies from the second and third terms do not alter the physics, i.e., the energy is again minimized for $\{{\cred\tilde{a}}, {\cgr\tilde{b}}, {\cbl\tilde{c}}\}$ along the cubic axes.
Finally, there is a coupling between two $\eta$-types, $J_{{\cred A}{\cgr B}}({\cred\eta_1\eta_6})({\cgr\eta_2\eta_5})$, where $J_{{\cred A}{\cgr B}}\!=\!\frac{7|\Gamma|}{384}{\cred\tilde{a}}^2{\cgr\tilde{b}}^2$, etc.  These terms favor also $\{{\cred\tilde{a}}, {\cgr\tilde{b}}, {\cbl\tilde{c}}\}$ along the cubic axes. 
Altogether then, the leading, second-order terms give an excellent description of the order-by-disorder physics.

{\cdbl {\it Role of perturbations --}}  
Classically, the ground state degeneracy of the $\Gamma$ model is immediately unstable against other terms in the Hamiltonian, such as NN or next-NN Kitaev coupling, $K_1$ and $K_2$.~\cite{IoannisLong} Quantum-mechanically, however, the physics of the $\Gamma$ model survives in a finite region of parameter space, where the order-by-disorder energy scale ($\propto\Gamma S$) outweighs the classical energy contributions from the other terms.
The extent of this region depends on the specific JKK system and the nature of the perturbations. 
We can foresee, however, that the highly frustrated physics of the positive $\Gamma$ model should be more stable on the ferromagnet $K_1$ side, because $K_1$ acts to renormalize $J_{{\cred A}}$, $J_{{\cgr B}}$ and $J_{{\cbl C}}$ by $-K_1S^2{\cred \tilde{a}}^2$, $-K_1S^2{\cgr \tilde{b}}^2$ and $-K_1S^2{\cbl \tilde{c}}^2$, respectively. This is important because $K_1$ is FM in all JKK materials.

{\cdbl {\it Discussion --}} 
Our predictions are consistent with the XMCD data in $\beta$-Li$_2$IrO$_3$,~\cite{Takayama2015} for either sign of $\Gamma$. 
At ambient pressure, $\beta$-Li$_2$IrO$_3$ shows an incommensurate magnetic order which is very close to a partially polarized state.~\cite{Takayama2015} The {\it ab initio} studies~\cite{Kim2016,Liviu2016} show that $|\Gamma|$, which is already appreciable at ambient pressure, increases by 10-15 $\%$ at 2 GPa, while $|K_1|$ drops by a remarkable 40-50 $\%$. 
Clearly then, the system departs very quickly from the vicinity of the partially polarized state, toward the classical manifold of the $\Gamma$ model, and eventually orders either at a very small energy scale if $\Gamma\!>\!0$, or at the scale $E_{\text{ani}}$ if $\Gamma\!<\!0$. Either way, one expects a strong suppression of the field-induced ferromagnetic moments. 
According to {\it ab initio} studies~\cite{Kim2016,Liviu2016} and fits to experiments,~\cite{Lee2016} $\Gamma$ is negative, which would mean that the system orders in the six-sublattice,  non-coplanar state discussed above. This prediction can be confirmed e.g., by local probes, such as NMR or $\mu$SR.

On a broader perspective, we have shown that Mott insulators with strong spin-orbit coupling and bond-dependent interactions host yet another exotic correlated regime, besides the well-known Kitaev QSL. This regime is governed by a classical spin liquid instability and unconventional spin-spin correlations along closed or open strings. Remarkably, the key predictions are common for all available 2D and 3D JKK materials, providing a distinct platform for further studies in this direction.

\acknowledgments
We thank A. Tsirlin, T. Takayama, R. Yadav, L. Hozoi, J. van den Brink, C. Batista, Y. Sizyuk, J. Reuther, R. Coldea, and Y. B. Kim for fruitful discussions, and acknowledge support from NSF Grant No. DMR-1511768. 
IR acknowledges the hospitality of MPI-PKS of Dresden where part of this work was done, and NP acknowledges the hospitality of Aspen Center for Physics and NSF GRANT No. PHY-1066293.


\begin{thebibliography}{61}%
\makeatletter
\providecommand \@ifxundefined [1]{%
 \@ifx{#1\undefined}
}%
\providecommand \@ifnum [1]{%
 \ifnum #1\expandafter \@firstoftwo
 \else \expandafter \@secondoftwo
 \fi
}%
\providecommand \@ifx [1]{%
 \ifx #1\expandafter \@firstoftwo
 \else \expandafter \@secondoftwo
 \fi
}%
\providecommand \natexlab [1]{#1}%
\providecommand \enquote  [1]{``#1''}%
\providecommand \bibnamefont  [1]{#1}%
\providecommand \bibfnamefont [1]{#1}%
\providecommand \citenamefont [1]{#1}%
\providecommand \href@noop [0]{\@secondoftwo}%
\providecommand \href [0]{\begingroup \@sanitize@url \@href}%
\providecommand \@href[1]{\@@startlink{#1}\@@href}%
\providecommand \@@href[1]{\endgroup#1\@@endlink}%
\providecommand \@sanitize@url [0]{\catcode `\\12\catcode `\$12\catcode
  `\&12\catcode `\#12\catcode `\^12\catcode `\_12\catcode `\%12\relax}%
\providecommand \@@startlink[1]{}%
\providecommand \@@endlink[0]{}%
\providecommand \url  [0]{\begingroup\@sanitize@url \@url }%
\providecommand \@url [1]{\endgroup\@href {#1}{\urlprefix }}%
\providecommand \urlprefix  [0]{URL }%
\providecommand \Eprint [0]{\href }%
\providecommand \doibase [0]{http://dx.doi.org/}%
\providecommand \selectlanguage [0]{\@gobble}%
\providecommand \bibinfo  [0]{\@secondoftwo}%
\providecommand \bibfield  [0]{\@secondoftwo}%
\providecommand \translation [1]{[#1]}%
\providecommand \BibitemOpen [0]{}%
\providecommand \bibitemStop [0]{}%
\providecommand \bibitemNoStop [0]{.\EOS\space}%
\providecommand \EOS [0]{\spacefactor3000\relax}%
\providecommand \BibitemShut  [1]{\csname bibitem#1\endcsname}%
\let\auto@bib@innerbib\@empty
\bibitem [{\citenamefont {Anderson}(1973)}]{Anderson1973}%
  \BibitemOpen
  \bibfield  {author} {\bibinfo {author} {\bibfnamefont {P.}~\bibnamefont
  {Anderson}},\ }\href {\doibase 10.1016/0025-5408(73)90167-0} {\bibfield
  {journal} {\bibinfo  {journal} {Materials Research Bulletin}\ }\textbf
  {\bibinfo {volume} {8}},\ \bibinfo {pages} {153 } (\bibinfo {year}
  {1973})}\BibitemShut {NoStop}%
\bibitem [{HFM(2011)}]{HFMBook}%
  \BibitemOpen
  \href@noop {} {\emph {\bibinfo {title} {Introduction to Frustrated Magnetism:
  Materials, Experiments, Theory}}}\ (\bibinfo  {publisher} {Springer Series in
  Solid-State Sciences},\ \bibinfo {address} {Berlin},\ \bibinfo {year}
  {2011})\BibitemShut {NoStop}%
\bibitem [{\citenamefont {Balents}(2010)}]{Balents2010}%
  \BibitemOpen
  \bibfield  {author} {\bibinfo {author} {\bibfnamefont {L.}~\bibnamefont
  {Balents}},\ }\href {\doibase 10.1038/nature08917} {\bibfield  {journal}
  {\bibinfo  {journal} {Nature}\ }\textbf {\bibinfo {volume} {464}},\ \bibinfo
  {pages} {199} (\bibinfo {year} {2010})}\BibitemShut {NoStop}%
\bibitem [{\citenamefont {Savary}\ and\ \citenamefont
  {Balents}(2016)}]{Savary2016}%
  \BibitemOpen
  \bibfield  {author} {\bibinfo {author} {\bibfnamefont {L.}~\bibnamefont
  {Savary}}\ and\ \bibinfo {author} {\bibfnamefont {L.}~\bibnamefont
  {Balents}},\ }\href {http://arxiv.org/abs/1601.03742} {\bibfield  {journal}
  {\bibinfo  {journal} {arXiv:1601.03742}\ } (\bibinfo {year}
  {2016})}\BibitemShut {NoStop}%
\bibitem [{\citenamefont {Kitaev}(2006)}]{Kitaev2006}%
  \BibitemOpen
  \bibfield  {author} {\bibinfo {author} {\bibfnamefont {A.}~\bibnamefont
  {Kitaev}},\ }\href {\doibase http://dx.doi.org/10.1016/j.aop.2005.10.005}
  {\bibfield  {journal} {\bibinfo  {journal} {Annals of Physics}\ }\textbf
  {\bibinfo {volume} {321}},\ \bibinfo {pages} {2 } (\bibinfo {year}
  {2006})}\BibitemShut {NoStop}%
\bibitem [{\citenamefont {Jackeli}\ and\ \citenamefont
  {Khaliullin}(2009)}]{Jackeli2009}%
  \BibitemOpen
  \bibfield  {author} {\bibinfo {author} {\bibfnamefont {G.}~\bibnamefont
  {Jackeli}}\ and\ \bibinfo {author} {\bibfnamefont {G.}~\bibnamefont
  {Khaliullin}},\ }\href {\doibase 10.1103/PhysRevLett.102.017205} {\bibfield
  {journal} {\bibinfo  {journal} {Phys. Rev. Lett.}\ }\textbf {\bibinfo
  {volume} {102}},\ \bibinfo {pages} {017205} (\bibinfo {year}
  {2009})}\BibitemShut {NoStop}%
\bibitem [{\citenamefont {Chaloupka}\ \emph {et~al.}(2010)\citenamefont
  {Chaloupka}, \citenamefont {Jackeli},\ and\ \citenamefont
  {Khaliullin}}]{Jackeli2010}%
  \BibitemOpen
  \bibfield  {author} {\bibinfo {author} {\bibfnamefont {J.}~\bibnamefont
  {Chaloupka}}, \bibinfo {author} {\bibfnamefont {G.}~\bibnamefont {Jackeli}},
  \ and\ \bibinfo {author} {\bibfnamefont {G.}~\bibnamefont {Khaliullin}},\
  }\href {\doibase 10.1103/PhysRevLett.105.027204} {\bibfield  {journal}
  {\bibinfo  {journal} {Phys. Rev. Lett.}\ }\textbf {\bibinfo {volume} {105}},\
  \bibinfo {pages} {027204} (\bibinfo {year} {2010})}\BibitemShut {NoStop}%
\bibitem [{\citenamefont {Chaloupka}\ \emph {et~al.}(2013)\citenamefont
  {Chaloupka}, \citenamefont {Jackeli},\ and\ \citenamefont
  {Khaliullin}}]{Jackeli2013}%
  \BibitemOpen
  \bibfield  {author} {\bibinfo {author} {\bibfnamefont {J.}~\bibnamefont
  {Chaloupka}}, \bibinfo {author} {\bibfnamefont {G.}~\bibnamefont {Jackeli}},
  \ and\ \bibinfo {author} {\bibfnamefont {G.}~\bibnamefont {Khaliullin}},\
  }\href {\doibase 10.1103/PhysRevLett.110.097204} {\bibfield  {journal}
  {\bibinfo  {journal} {Phys. Rev. Lett.}\ }\textbf {\bibinfo {volume} {110}},\
  \bibinfo {pages} {097204} (\bibinfo {year} {2013})}\BibitemShut {NoStop}%
\bibitem [{\citenamefont {Mandal}\ and\ \citenamefont
  {Surendran}(2009)}]{Mandal2009}%
  \BibitemOpen
  \bibfield  {author} {\bibinfo {author} {\bibfnamefont {S.}~\bibnamefont
  {Mandal}}\ and\ \bibinfo {author} {\bibfnamefont {N.}~\bibnamefont
  {Surendran}},\ }\href {\doibase 10.1103/PhysRevB.79.024426} {\bibfield
  {journal} {\bibinfo  {journal} {Phys. Rev. B}\ }\textbf {\bibinfo {volume}
  {79}},\ \bibinfo {pages} {024426} (\bibinfo {year} {2009})}\BibitemShut
  {NoStop}%
\bibitem [{\citenamefont {O'Brien}\ \emph {et~al.}(2016)\citenamefont
  {O'Brien}, \citenamefont {Hermanns},\ and\ \citenamefont
  {Trebst}}]{Obrien2016}%
  \BibitemOpen
  \bibfield  {author} {\bibinfo {author} {\bibfnamefont {K.}~\bibnamefont
  {O'Brien}}, \bibinfo {author} {\bibfnamefont {M.}~\bibnamefont {Hermanns}}, \
  and\ \bibinfo {author} {\bibfnamefont {S.}~\bibnamefont {Trebst}},\ }\href
  {\doibase 10.1103/PhysRevB.93.085101} {\bibfield  {journal} {\bibinfo
  {journal} {Phys. Rev. B}\ }\textbf {\bibinfo {volume} {93}},\ \bibinfo
  {pages} {085101} (\bibinfo {year} {2016})}\BibitemShut {NoStop}%
\bibitem [{\citenamefont {Witczak-Krempa}\ \emph {et~al.}(2014)\citenamefont
  {Witczak-Krempa}, \citenamefont {Chen}, \citenamefont {Kim},\ and\
  \citenamefont {Balents}}]{Krempa2014}%
  \BibitemOpen
  \bibfield  {author} {\bibinfo {author} {\bibfnamefont {W.}~\bibnamefont
  {Witczak-Krempa}}, \bibinfo {author} {\bibfnamefont {G.}~\bibnamefont
  {Chen}}, \bibinfo {author} {\bibfnamefont {Y.~B.}\ \bibnamefont {Kim}}, \
  and\ \bibinfo {author} {\bibfnamefont {L.}~\bibnamefont {Balents}},\ }\href
  {\doibase 10.1146/annurev-conmatphys-020911-125138} {\bibfield  {journal}
  {\bibinfo  {journal} {Annual Review of Condensed Matter Physics}\ }\textbf
  {\bibinfo {volume} {5}},\ \bibinfo {pages} {57} (\bibinfo {year}
  {2014})}\BibitemShut {NoStop}%
\bibitem [{\citenamefont {Singh}\ and\ \citenamefont
  {Gegenwart}(2010)}]{Singh2010}%
  \BibitemOpen
  \bibfield  {author} {\bibinfo {author} {\bibfnamefont {Y.}~\bibnamefont
  {Singh}}\ and\ \bibinfo {author} {\bibfnamefont {P.}~\bibnamefont
  {Gegenwart}},\ }\href {\doibase 10.1103/PhysRevB.82.064412} {\bibfield
  {journal} {\bibinfo  {journal} {Phys. Rev. B}\ }\textbf {\bibinfo {volume}
  {82}},\ \bibinfo {pages} {064412} (\bibinfo {year} {2010})}\BibitemShut
  {NoStop}%
\bibitem [{\citenamefont {Singh}\ \emph {et~al.}(2012)\citenamefont {Singh},
  \citenamefont {Manni}, \citenamefont {Reuther}, \citenamefont {Berlijn},
  \citenamefont {Thomale}, \citenamefont {Ku}, \citenamefont {Trebst},\ and\
  \citenamefont {Gegenwart}}]{Singh2012}%
  \BibitemOpen
  \bibfield  {author} {\bibinfo {author} {\bibfnamefont {Y.}~\bibnamefont
  {Singh}}, \bibinfo {author} {\bibfnamefont {S.}~\bibnamefont {Manni}},
  \bibinfo {author} {\bibfnamefont {J.}~\bibnamefont {Reuther}}, \bibinfo
  {author} {\bibfnamefont {T.}~\bibnamefont {Berlijn}}, \bibinfo {author}
  {\bibfnamefont {R.}~\bibnamefont {Thomale}}, \bibinfo {author} {\bibfnamefont
  {W.}~\bibnamefont {Ku}}, \bibinfo {author} {\bibfnamefont {S.}~\bibnamefont
  {Trebst}}, \ and\ \bibinfo {author} {\bibfnamefont {P.}~\bibnamefont
  {Gegenwart}},\ }\href {\doibase 10.1103/PhysRevLett.108.127203} {\bibfield
  {journal} {\bibinfo  {journal} {Phys. Rev. Lett.}\ }\textbf {\bibinfo
  {volume} {108}},\ \bibinfo {pages} {127203} (\bibinfo {year}
  {2012})}\BibitemShut {NoStop}%
\bibitem [{\citenamefont {Liu}\ \emph {et~al.}(2011)\citenamefont {Liu},
  \citenamefont {Berlijn}, \citenamefont {Yin}, \citenamefont {Ku},
  \citenamefont {Tsvelik}, \citenamefont {Kim}, \citenamefont {Gretarsson},
  \citenamefont {Singh}, \citenamefont {Gegenwart},\ and\ \citenamefont
  {Hill}}]{Liu2011}%
  \BibitemOpen
  \bibfield  {author} {\bibinfo {author} {\bibfnamefont {X.}~\bibnamefont
  {Liu}}, \bibinfo {author} {\bibfnamefont {T.}~\bibnamefont {Berlijn}},
  \bibinfo {author} {\bibfnamefont {W.-G.}\ \bibnamefont {Yin}}, \bibinfo
  {author} {\bibfnamefont {W.}~\bibnamefont {Ku}}, \bibinfo {author}
  {\bibfnamefont {A.}~\bibnamefont {Tsvelik}}, \bibinfo {author} {\bibfnamefont
  {Y.-J.}\ \bibnamefont {Kim}}, \bibinfo {author} {\bibfnamefont
  {H.}~\bibnamefont {Gretarsson}}, \bibinfo {author} {\bibfnamefont
  {Y.}~\bibnamefont {Singh}}, \bibinfo {author} {\bibfnamefont
  {P.}~\bibnamefont {Gegenwart}}, \ and\ \bibinfo {author} {\bibfnamefont
  {J.~P.}\ \bibnamefont {Hill}},\ }\href {\doibase 10.1103/PhysRevB.83.220403}
  {\bibfield  {journal} {\bibinfo  {journal} {Phys. Rev. B}\ }\textbf {\bibinfo
  {volume} {83}},\ \bibinfo {pages} {220403} (\bibinfo {year}
  {2011})}\BibitemShut {NoStop}%
\bibitem [{\citenamefont {Choi}\ \emph {et~al.}(2012)\citenamefont {Choi},
  \citenamefont {Coldea}, \citenamefont {Kolmogorov}, \citenamefont
  {Lancaster}, \citenamefont {Mazin}, \citenamefont {Blundell}, \citenamefont
  {Radaelli}, \citenamefont {Singh}, \citenamefont {Gegenwart}, \citenamefont
  {Choi}, \citenamefont {Cheong}, \citenamefont {Baker}, \citenamefont
  {Stock},\ and\ \citenamefont {Taylor}}]{Choi2012}%
  \BibitemOpen
  \bibfield  {author} {\bibinfo {author} {\bibfnamefont {S.~K.}\ \bibnamefont
  {Choi}}, \bibinfo {author} {\bibfnamefont {R.}~\bibnamefont {Coldea}},
  \bibinfo {author} {\bibfnamefont {A.~N.}\ \bibnamefont {Kolmogorov}},
  \bibinfo {author} {\bibfnamefont {T.}~\bibnamefont {Lancaster}}, \bibinfo
  {author} {\bibfnamefont {I.~I.}\ \bibnamefont {Mazin}}, \bibinfo {author}
  {\bibfnamefont {S.~J.}\ \bibnamefont {Blundell}}, \bibinfo {author}
  {\bibfnamefont {P.~G.}\ \bibnamefont {Radaelli}}, \bibinfo {author}
  {\bibfnamefont {Y.}~\bibnamefont {Singh}}, \bibinfo {author} {\bibfnamefont
  {P.}~\bibnamefont {Gegenwart}}, \bibinfo {author} {\bibfnamefont {K.~R.}\
  \bibnamefont {Choi}}, \bibinfo {author} {\bibfnamefont {S.-W.}\ \bibnamefont
  {Cheong}}, \bibinfo {author} {\bibfnamefont {P.~J.}\ \bibnamefont {Baker}},
  \bibinfo {author} {\bibfnamefont {C.}~\bibnamefont {Stock}}, \ and\ \bibinfo
  {author} {\bibfnamefont {J.}~\bibnamefont {Taylor}},\ }\href {\doibase
  10.1103/PhysRevLett.108.127204} {\bibfield  {journal} {\bibinfo  {journal}
  {Phys. Rev. Lett.}\ }\textbf {\bibinfo {volume} {108}},\ \bibinfo {pages}
  {127204} (\bibinfo {year} {2012})}\BibitemShut {NoStop}%
\bibitem [{\citenamefont {Ye}\ \emph {et~al.}(2012)\citenamefont {Ye},
  \citenamefont {Chi}, \citenamefont {Cao}, \citenamefont {Chakoumakos},
  \citenamefont {Fernandez-Baca}, \citenamefont {Custelcean}, \citenamefont
  {Qi}, \citenamefont {Korneta},\ and\ \citenamefont {Cao}}]{Ye2012}%
  \BibitemOpen
  \bibfield  {author} {\bibinfo {author} {\bibfnamefont {F.}~\bibnamefont
  {Ye}}, \bibinfo {author} {\bibfnamefont {S.}~\bibnamefont {Chi}}, \bibinfo
  {author} {\bibfnamefont {H.}~\bibnamefont {Cao}}, \bibinfo {author}
  {\bibfnamefont {B.~C.}\ \bibnamefont {Chakoumakos}}, \bibinfo {author}
  {\bibfnamefont {J.~A.}\ \bibnamefont {Fernandez-Baca}}, \bibinfo {author}
  {\bibfnamefont {R.}~\bibnamefont {Custelcean}}, \bibinfo {author}
  {\bibfnamefont {T.~F.}\ \bibnamefont {Qi}}, \bibinfo {author} {\bibfnamefont
  {O.~B.}\ \bibnamefont {Korneta}}, \ and\ \bibinfo {author} {\bibfnamefont
  {G.}~\bibnamefont {Cao}},\ }\href
  {http://link.aps.org/doi/10.1103/PhysRevB.85.180403} {\bibfield  {journal}
  {\bibinfo  {journal} {Phys. Rev. B}\ }\textbf {\bibinfo {volume} {85}},\
  \bibinfo {pages} {180403} (\bibinfo {year} {2012})}\BibitemShut {NoStop}%
\bibitem [{\citenamefont {Hwan~Chun}\ \emph {et~al.}(2015)\citenamefont
  {Hwan~Chun}, \citenamefont {Kim}, \citenamefont {Kim}, \citenamefont {Zheng},
  \citenamefont {Stoumpos}, \citenamefont {Malliakas}, \citenamefont
  {Mitchell}, \citenamefont {Mehlawat}, , \citenamefont {Singh}, \citenamefont
  {Choi}, \citenamefont {Gog}, \citenamefont {Al-Zein}, \citenamefont {Sala},
  \citenamefont {Krisch}, \citenamefont {Chaloupka}, \citenamefont {Jackeli},
  \citenamefont {Khaliullin},\ and\ \citenamefont {Kim}}]{Chun2015}%
  \BibitemOpen
  \bibfield  {author} {\bibinfo {author} {\bibfnamefont {S.}~\bibnamefont
  {Hwan~Chun}}, \bibinfo {author} {\bibfnamefont {J.-W.}\ \bibnamefont {Kim}},
  \bibinfo {author} {\bibfnamefont {J.}~\bibnamefont {Kim}}, \bibinfo {author}
  {\bibfnamefont {H.}~\bibnamefont {Zheng}}, \bibinfo {author} {\bibfnamefont
  {C.~C.}\ \bibnamefont {Stoumpos}}, \bibinfo {author} {\bibfnamefont {C.~D.}\
  \bibnamefont {Malliakas}}, \bibinfo {author} {\bibfnamefont {J.~F.}\
  \bibnamefont {Mitchell}}, \bibinfo {author} {\bibfnamefont {K.}~\bibnamefont
  {Mehlawat}}, , \bibinfo {author} {\bibfnamefont {Y.}~\bibnamefont {Singh}},
  \bibinfo {author} {\bibfnamefont {Y.}~\bibnamefont {Choi}}, \bibinfo {author}
  {\bibfnamefont {T.}~\bibnamefont {Gog}}, \bibinfo {author} {\bibfnamefont
  {A.}~\bibnamefont {Al-Zein}}, \bibinfo {author} {\bibfnamefont {M.~M.}\
  \bibnamefont {Sala}}, \bibinfo {author} {\bibfnamefont {M.}~\bibnamefont
  {Krisch}}, \bibinfo {author} {\bibfnamefont {J.}~\bibnamefont {Chaloupka}},
  \bibinfo {author} {\bibfnamefont {G.}~\bibnamefont {Jackeli}}, \bibinfo
  {author} {\bibfnamefont {G.}~\bibnamefont {Khaliullin}}, \ and\ \bibinfo
  {author} {\bibfnamefont {B.~J.}\ \bibnamefont {Kim}},\ }\href {\doibase
  10.1038/nphys3322} {\bibfield  {journal} {\bibinfo  {journal} {Nat Phys}\
  }\textbf {\bibinfo {volume} {10}},\ \bibinfo {pages} {1038} (\bibinfo {year}
  {2015})}\BibitemShut {NoStop}%
\bibitem [{\citenamefont {Williams}\ \emph {et~al.}(2016)\citenamefont
  {Williams}, \citenamefont {Johnson}, \citenamefont {Freund}, \citenamefont
  {Choi}, \citenamefont {Jesche}, \citenamefont {Kimchi}, \citenamefont
  {Manni}, \citenamefont {Bombardi}, \citenamefont {Manuel}, \citenamefont
  {Gegenwart},\ and\ \citenamefont {Coldea}}]{Williams2016}%
  \BibitemOpen
  \bibfield  {author} {\bibinfo {author} {\bibfnamefont {S.~C.}\ \bibnamefont
  {Williams}}, \bibinfo {author} {\bibfnamefont {R.~D.}\ \bibnamefont
  {Johnson}}, \bibinfo {author} {\bibfnamefont {F.}~\bibnamefont {Freund}},
  \bibinfo {author} {\bibfnamefont {S.}~\bibnamefont {Choi}}, \bibinfo {author}
  {\bibfnamefont {A.}~\bibnamefont {Jesche}}, \bibinfo {author} {\bibfnamefont
  {I.}~\bibnamefont {Kimchi}}, \bibinfo {author} {\bibfnamefont
  {S.}~\bibnamefont {Manni}}, \bibinfo {author} {\bibfnamefont
  {A.}~\bibnamefont {Bombardi}}, \bibinfo {author} {\bibfnamefont
  {P.}~\bibnamefont {Manuel}}, \bibinfo {author} {\bibfnamefont
  {P.}~\bibnamefont {Gegenwart}}, \ and\ \bibinfo {author} {\bibfnamefont
  {R.}~\bibnamefont {Coldea}},\ }\href {\doibase 10.1103/PhysRevB.93.195158}
  {\bibfield  {journal} {\bibinfo  {journal} {Phys. Rev. B}\ }\textbf {\bibinfo
  {volume} {93}},\ \bibinfo {pages} {195158} (\bibinfo {year}
  {2016})}\BibitemShut {NoStop}%
\bibitem [{\citenamefont {Plumb}\ \emph {et~al.}(2014)\citenamefont {Plumb},
  \citenamefont {Clancy}, \citenamefont {Sandilands}, \citenamefont {Shankar},
  \citenamefont {Hu}, \citenamefont {Burch}, \citenamefont {Kee},\ and\
  \citenamefont {Kim}}]{Plumb2014}%
  \BibitemOpen
  \bibfield  {author} {\bibinfo {author} {\bibfnamefont {K.~W.}\ \bibnamefont
  {Plumb}}, \bibinfo {author} {\bibfnamefont {J.~P.}\ \bibnamefont {Clancy}},
  \bibinfo {author} {\bibfnamefont {L.~J.}\ \bibnamefont {Sandilands}},
  \bibinfo {author} {\bibfnamefont {V.~V.}\ \bibnamefont {Shankar}}, \bibinfo
  {author} {\bibfnamefont {Y.~F.}\ \bibnamefont {Hu}}, \bibinfo {author}
  {\bibfnamefont {K.~S.}\ \bibnamefont {Burch}}, \bibinfo {author}
  {\bibfnamefont {H.-Y.}\ \bibnamefont {Kee}}, \ and\ \bibinfo {author}
  {\bibfnamefont {Y.-J.}\ \bibnamefont {Kim}},\ }\href {\doibase
  10.1103/PhysRevB.90.041112} {\bibfield  {journal} {\bibinfo  {journal} {Phys.
  Rev. B}\ }\textbf {\bibinfo {volume} {90}},\ \bibinfo {pages} {041112}
  (\bibinfo {year} {2014})}\BibitemShut {NoStop}%
\bibitem [{\citenamefont {Sears}\ \emph {et~al.}(2015)\citenamefont {Sears},
  \citenamefont {Songvilay}, \citenamefont {Plumb}, \citenamefont {Clancy},
  \citenamefont {Qiu}, \citenamefont {Zhao}, \citenamefont {Parshall},\ and\
  \citenamefont {Kim}}]{Sears2015}%
  \BibitemOpen
  \bibfield  {author} {\bibinfo {author} {\bibfnamefont {J.~A.}\ \bibnamefont
  {Sears}}, \bibinfo {author} {\bibfnamefont {M.}~\bibnamefont {Songvilay}},
  \bibinfo {author} {\bibfnamefont {K.~W.}\ \bibnamefont {Plumb}}, \bibinfo
  {author} {\bibfnamefont {J.~P.}\ \bibnamefont {Clancy}}, \bibinfo {author}
  {\bibfnamefont {Y.}~\bibnamefont {Qiu}}, \bibinfo {author} {\bibfnamefont
  {Y.}~\bibnamefont {Zhao}}, \bibinfo {author} {\bibfnamefont {D.}~\bibnamefont
  {Parshall}}, \ and\ \bibinfo {author} {\bibfnamefont {Y.-J.}\ \bibnamefont
  {Kim}},\ }\href {\doibase 10.1103/PhysRevB.91.144420} {\bibfield  {journal}
  {\bibinfo  {journal} {Phys. Rev. B}\ }\textbf {\bibinfo {volume} {91}},\
  \bibinfo {pages} {144420} (\bibinfo {year} {2015})}\BibitemShut {NoStop}%
\bibitem [{\citenamefont {Kubota}\ \emph {et~al.}(2015)\citenamefont {Kubota},
  \citenamefont {Tanaka}, \citenamefont {Ono}, \citenamefont {Narumi},\ and\
  \citenamefont {Kindo}}]{Kubota2015}%
  \BibitemOpen
  \bibfield  {author} {\bibinfo {author} {\bibfnamefont {Y.}~\bibnamefont
  {Kubota}}, \bibinfo {author} {\bibfnamefont {H.}~\bibnamefont {Tanaka}},
  \bibinfo {author} {\bibfnamefont {T.}~\bibnamefont {Ono}}, \bibinfo {author}
  {\bibfnamefont {Y.}~\bibnamefont {Narumi}}, \ and\ \bibinfo {author}
  {\bibfnamefont {K.}~\bibnamefont {Kindo}},\ }\href {\doibase
  10.1103/PhysRevB.91.094422} {\bibfield  {journal} {\bibinfo  {journal} {Phys.
  Rev. B}\ }\textbf {\bibinfo {volume} {91}},\ \bibinfo {pages} {094422}
  (\bibinfo {year} {2015})}\BibitemShut {NoStop}%
\bibitem [{\citenamefont {Majumder}\ \emph {et~al.}(2015)\citenamefont
  {Majumder}, \citenamefont {Schmidt}, \citenamefont {Rosner}, \citenamefont
  {Tsirlin}, \citenamefont {Yasuoka},\ and\ \citenamefont
  {Baenitz}}]{Majumder2015}%
  \BibitemOpen
  \bibfield  {author} {\bibinfo {author} {\bibfnamefont {M.}~\bibnamefont
  {Majumder}}, \bibinfo {author} {\bibfnamefont {M.}~\bibnamefont {Schmidt}},
  \bibinfo {author} {\bibfnamefont {H.}~\bibnamefont {Rosner}}, \bibinfo
  {author} {\bibfnamefont {A.~A.}\ \bibnamefont {Tsirlin}}, \bibinfo {author}
  {\bibfnamefont {H.}~\bibnamefont {Yasuoka}}, \ and\ \bibinfo {author}
  {\bibfnamefont {M.}~\bibnamefont {Baenitz}},\ }\href {\doibase
  10.1103/PhysRevB.91.180401} {\bibfield  {journal} {\bibinfo  {journal} {Phys.
  Rev. B}\ }\textbf {\bibinfo {volume} {91}},\ \bibinfo {pages} {180401}
  (\bibinfo {year} {2015})}\BibitemShut {NoStop}%
\bibitem [{\citenamefont {Banerjee}\ \emph {et~al.}(2016)\citenamefont
  {Banerjee}, \citenamefont {Bridges}, \citenamefont {Yan}, \citenamefont
  {Aczel}, \citenamefont {Li}, \citenamefont {Stone}, \citenamefont {Granroth},
  \citenamefont {Lumsden}, \citenamefont {Yiu}, \citenamefont {Knolle} \emph
  {et~al.}}]{Banerjee2016}%
  \BibitemOpen
  \bibfield  {author} {\bibinfo {author} {\bibfnamefont {A.}~\bibnamefont
  {Banerjee}}, \bibinfo {author} {\bibfnamefont {C.}~\bibnamefont {Bridges}},
  \bibinfo {author} {\bibfnamefont {J.-Q.}\ \bibnamefont {Yan}}, \bibinfo
  {author} {\bibfnamefont {A.}~\bibnamefont {Aczel}}, \bibinfo {author}
  {\bibfnamefont {L.}~\bibnamefont {Li}}, \bibinfo {author} {\bibfnamefont
  {M.}~\bibnamefont {Stone}}, \bibinfo {author} {\bibfnamefont
  {G.}~\bibnamefont {Granroth}}, \bibinfo {author} {\bibfnamefont
  {M.}~\bibnamefont {Lumsden}}, \bibinfo {author} {\bibfnamefont
  {Y.}~\bibnamefont {Yiu}}, \bibinfo {author} {\bibfnamefont {J.}~\bibnamefont
  {Knolle}},  \emph {et~al.},\ }\href {\doibase 10.1038/nmat4604} {\bibfield
  {journal} {\bibinfo  {journal} {Nature materials}\ } (\bibinfo {year}
  {2016}),\ 10.1038/nmat4604}\BibitemShut {NoStop}%
\bibitem [{\citenamefont {Johnson}\ \emph {et~al.}(2015)\citenamefont
  {Johnson}, \citenamefont {Williams}, \citenamefont {Haghighirad},
  \citenamefont {Singleton}, \citenamefont {Zapf}, \citenamefont {Manuel},
  \citenamefont {Mazin}, \citenamefont {Li}, \citenamefont {Jeschke},
  \citenamefont {Valent\'{\i}},\ and\ \citenamefont {Coldea}}]{Johnson2015}%
  \BibitemOpen
  \bibfield  {author} {\bibinfo {author} {\bibfnamefont {R.~D.}\ \bibnamefont
  {Johnson}}, \bibinfo {author} {\bibfnamefont {S.~C.}\ \bibnamefont
  {Williams}}, \bibinfo {author} {\bibfnamefont {A.~A.}\ \bibnamefont
  {Haghighirad}}, \bibinfo {author} {\bibfnamefont {J.}~\bibnamefont
  {Singleton}}, \bibinfo {author} {\bibfnamefont {V.}~\bibnamefont {Zapf}},
  \bibinfo {author} {\bibfnamefont {P.}~\bibnamefont {Manuel}}, \bibinfo
  {author} {\bibfnamefont {I.~I.}\ \bibnamefont {Mazin}}, \bibinfo {author}
  {\bibfnamefont {Y.}~\bibnamefont {Li}}, \bibinfo {author} {\bibfnamefont
  {H.~O.}\ \bibnamefont {Jeschke}}, \bibinfo {author} {\bibfnamefont
  {R.}~\bibnamefont {Valent\'{\i}}}, \ and\ \bibinfo {author} {\bibfnamefont
  {R.}~\bibnamefont {Coldea}},\ }\href {\doibase 10.1103/PhysRevB.92.235119}
  {\bibfield  {journal} {\bibinfo  {journal} {Phys. Rev. B}\ }\textbf {\bibinfo
  {volume} {92}},\ \bibinfo {pages} {235119} (\bibinfo {year}
  {2015})}\BibitemShut {NoStop}%
\bibitem [{\citenamefont {Biffin}\ \emph
  {et~al.}(2014{\natexlab{a}})\citenamefont {Biffin}, \citenamefont {Johnson},
  \citenamefont {Choi}, \citenamefont {Freund}, \citenamefont {Manni},
  \citenamefont {Bombardi}, \citenamefont {Manuel}, \citenamefont {Gegenwart},\
  and\ \citenamefont {Coldea}}]{Biffin2014a}%
  \BibitemOpen
  \bibfield  {author} {\bibinfo {author} {\bibfnamefont {A.}~\bibnamefont
  {Biffin}}, \bibinfo {author} {\bibfnamefont {R.~D.}\ \bibnamefont {Johnson}},
  \bibinfo {author} {\bibfnamefont {S.}~\bibnamefont {Choi}}, \bibinfo {author}
  {\bibfnamefont {F.}~\bibnamefont {Freund}}, \bibinfo {author} {\bibfnamefont
  {S.}~\bibnamefont {Manni}}, \bibinfo {author} {\bibfnamefont
  {A.}~\bibnamefont {Bombardi}}, \bibinfo {author} {\bibfnamefont
  {P.}~\bibnamefont {Manuel}}, \bibinfo {author} {\bibfnamefont
  {P.}~\bibnamefont {Gegenwart}}, \ and\ \bibinfo {author} {\bibfnamefont
  {R.}~\bibnamefont {Coldea}},\ }\href {\doibase 10.1103/PhysRevB.90.205116}
  {\bibfield  {journal} {\bibinfo  {journal} {Phys. Rev. B}\ }\textbf {\bibinfo
  {volume} {90}},\ \bibinfo {pages} {205116} (\bibinfo {year}
  {2014}{\natexlab{a}})}\BibitemShut {NoStop}%
\bibitem [{\citenamefont {Biffin}\ \emph
  {et~al.}(2014{\natexlab{b}})\citenamefont {Biffin}, \citenamefont {Johnson},
  \citenamefont {Kimchi}, \citenamefont {Morris}, \citenamefont {Bombardi},
  \citenamefont {Analytis}, \citenamefont {Vishwanath},\ and\ \citenamefont
  {Coldea}}]{Biffin2014b}%
  \BibitemOpen
  \bibfield  {author} {\bibinfo {author} {\bibfnamefont {A.}~\bibnamefont
  {Biffin}}, \bibinfo {author} {\bibfnamefont {R.~D.}\ \bibnamefont {Johnson}},
  \bibinfo {author} {\bibfnamefont {I.}~\bibnamefont {Kimchi}}, \bibinfo
  {author} {\bibfnamefont {R.}~\bibnamefont {Morris}}, \bibinfo {author}
  {\bibfnamefont {A.}~\bibnamefont {Bombardi}}, \bibinfo {author}
  {\bibfnamefont {J.~G.}\ \bibnamefont {Analytis}}, \bibinfo {author}
  {\bibfnamefont {A.}~\bibnamefont {Vishwanath}}, \ and\ \bibinfo {author}
  {\bibfnamefont {R.}~\bibnamefont {Coldea}},\ }\href {\doibase
  10.1103/PhysRevLett.113.197201} {\bibfield  {journal} {\bibinfo  {journal}
  {Phys. Rev. Lett.}\ }\textbf {\bibinfo {volume} {113}},\ \bibinfo {pages}
  {197201} (\bibinfo {year} {2014}{\natexlab{b}})}\BibitemShut {NoStop}%
\bibitem [{\citenamefont {Modic}\ \emph {et~al.}(2014)\citenamefont {Modic},
  \citenamefont {Smidt}, \citenamefont {Kimchi}, \citenamefont {Breznay},
  \citenamefont {Biffin}, \citenamefont {Choi}, \citenamefont {Johnson},
  \citenamefont {Coldea}, \citenamefont {Watkins-Curry}, \citenamefont
  {McCandess} \emph {et~al.}}]{Modic2014}%
  \BibitemOpen
  \bibfield  {author} {\bibinfo {author} {\bibfnamefont {K.}~\bibnamefont
  {Modic}}, \bibinfo {author} {\bibfnamefont {T.~E.}\ \bibnamefont {Smidt}},
  \bibinfo {author} {\bibfnamefont {I.}~\bibnamefont {Kimchi}}, \bibinfo
  {author} {\bibfnamefont {N.~P.}\ \bibnamefont {Breznay}}, \bibinfo {author}
  {\bibfnamefont {A.}~\bibnamefont {Biffin}}, \bibinfo {author} {\bibfnamefont
  {S.}~\bibnamefont {Choi}}, \bibinfo {author} {\bibfnamefont {R.~D.}\
  \bibnamefont {Johnson}}, \bibinfo {author} {\bibfnamefont {R.}~\bibnamefont
  {Coldea}}, \bibinfo {author} {\bibfnamefont {P.}~\bibnamefont
  {Watkins-Curry}}, \bibinfo {author} {\bibfnamefont {G.~T.}\ \bibnamefont
  {McCandess}},  \emph {et~al.},\ }\href {\doibase 10.1038/ncomms5203}
  {\bibfield  {journal} {\bibinfo  {journal} {Nature communications}\ }\textbf
  {\bibinfo {volume} {5}} (\bibinfo {year} {2014}),\
  10.1038/ncomms5203}\BibitemShut {NoStop}%
\bibitem [{\citenamefont {Takayama}\ \emph {et~al.}(2015)\citenamefont
  {Takayama}, \citenamefont {Kato}, \citenamefont {Dinnebier}, \citenamefont
  {Nuss}, \citenamefont {Kono}, \citenamefont {Veiga}, \citenamefont {Fabbris},
  \citenamefont {Haskel},\ and\ \citenamefont {Takagi}}]{Takayama2015}%
  \BibitemOpen
  \bibfield  {author} {\bibinfo {author} {\bibfnamefont {T.}~\bibnamefont
  {Takayama}}, \bibinfo {author} {\bibfnamefont {A.}~\bibnamefont {Kato}},
  \bibinfo {author} {\bibfnamefont {R.}~\bibnamefont {Dinnebier}}, \bibinfo
  {author} {\bibfnamefont {J.}~\bibnamefont {Nuss}}, \bibinfo {author}
  {\bibfnamefont {H.}~\bibnamefont {Kono}}, \bibinfo {author} {\bibfnamefont
  {L.~S.~I.}\ \bibnamefont {Veiga}}, \bibinfo {author} {\bibfnamefont
  {G.}~\bibnamefont {Fabbris}}, \bibinfo {author} {\bibfnamefont
  {D.}~\bibnamefont {Haskel}}, \ and\ \bibinfo {author} {\bibfnamefont
  {H.}~\bibnamefont {Takagi}},\ }\href {\doibase
  10.1103/PhysRevLett.114.077202} {\bibfield  {journal} {\bibinfo  {journal}
  {Phys. Rev. Lett.}\ }\textbf {\bibinfo {volume} {114}},\ \bibinfo {pages}
  {077202} (\bibinfo {year} {2015})}\BibitemShut {NoStop}%
\bibitem [{\citenamefont {Kimchi}\ \emph {et~al.}(2014)\citenamefont {Kimchi},
  \citenamefont {Analytis},\ and\ \citenamefont {Vishwanath}}]{Kimchi2014}%
  \BibitemOpen
  \bibfield  {author} {\bibinfo {author} {\bibfnamefont {I.}~\bibnamefont
  {Kimchi}}, \bibinfo {author} {\bibfnamefont {J.~G.}\ \bibnamefont
  {Analytis}}, \ and\ \bibinfo {author} {\bibfnamefont {A.}~\bibnamefont
  {Vishwanath}},\ }\href {\doibase 10.1103/PhysRevB.90.205126} {\bibfield
  {journal} {\bibinfo  {journal} {Phys. Rev. B}\ }\textbf {\bibinfo {volume}
  {90}},\ \bibinfo {pages} {205126} (\bibinfo {year} {2014})}\BibitemShut
  {NoStop}%
\bibitem [{\citenamefont {Katukuri}\ \emph {et~al.}(2014)\citenamefont
  {Katukuri}, \citenamefont {Nishimoto}, \citenamefont {Yushankhai},
  \citenamefont {Stoyanova}, \citenamefont {Kandpal}, \citenamefont {Choi},
  \citenamefont {Coldea}, \citenamefont {Rousochatzakis}, \citenamefont
  {Hozoi},\ and\ \citenamefont {van~den Brink}}]{Katukuri2014}%
  \BibitemOpen
  \bibfield  {author} {\bibinfo {author} {\bibfnamefont {V.~M.}\ \bibnamefont
  {Katukuri}}, \bibinfo {author} {\bibfnamefont {S.}~\bibnamefont {Nishimoto}},
  \bibinfo {author} {\bibfnamefont {V.}~\bibnamefont {Yushankhai}}, \bibinfo
  {author} {\bibfnamefont {A.}~\bibnamefont {Stoyanova}}, \bibinfo {author}
  {\bibfnamefont {H.}~\bibnamefont {Kandpal}}, \bibinfo {author} {\bibfnamefont
  {S.}~\bibnamefont {Choi}}, \bibinfo {author} {\bibfnamefont {R.}~\bibnamefont
  {Coldea}}, \bibinfo {author} {\bibfnamefont {I.}~\bibnamefont
  {Rousochatzakis}}, \bibinfo {author} {\bibfnamefont {L.}~\bibnamefont
  {Hozoi}}, \ and\ \bibinfo {author} {\bibfnamefont {J.}~\bibnamefont {van~den
  Brink}},\ }\href {http://stacks.iop.org/1367-2630/16/i=1/a=013056} {\bibfield
   {journal} {\bibinfo  {journal} {New J. Phys.}\ }\textbf {\bibinfo {volume}
  {16}},\ \bibinfo {pages} {013056} (\bibinfo {year} {2014})}\BibitemShut
  {NoStop}%
\bibitem [{\citenamefont {Sizyuk}\ \emph {et~al.}(2014)\citenamefont {Sizyuk},
  \citenamefont {Price}, \citenamefont {W\"olfle},\ and\ \citenamefont
  {Perkins}}]{Sizyuk2014}%
  \BibitemOpen
  \bibfield  {author} {\bibinfo {author} {\bibfnamefont {Y.}~\bibnamefont
  {Sizyuk}}, \bibinfo {author} {\bibfnamefont {C.}~\bibnamefont {Price}},
  \bibinfo {author} {\bibfnamefont {P.}~\bibnamefont {W\"olfle}}, \ and\
  \bibinfo {author} {\bibfnamefont {N.~B.}\ \bibnamefont {Perkins}},\ }\href
  {\doibase 10.1103/PhysRevB.90.155126} {\bibfield  {journal} {\bibinfo
  {journal} {Phys. Rev. B}\ }\textbf {\bibinfo {volume} {90}},\ \bibinfo
  {pages} {155126} (\bibinfo {year} {2014})}\BibitemShut {NoStop}%
\bibitem [{\citenamefont {Rau}\ \emph {et~al.}(2014)\citenamefont {Rau},
  \citenamefont {Lee},\ and\ \citenamefont {Kee}}]{Rau2014}%
  \BibitemOpen
  \bibfield  {author} {\bibinfo {author} {\bibfnamefont {J.~G.}\ \bibnamefont
  {Rau}}, \bibinfo {author} {\bibfnamefont {E.~K.-H.}\ \bibnamefont {Lee}}, \
  and\ \bibinfo {author} {\bibfnamefont {H.-Y.}\ \bibnamefont {Kee}},\ }\href
  {\doibase 10.1103/PhysRevLett.112.077204} {\bibfield  {journal} {\bibinfo
  {journal} {Phys. Rev. Lett.}\ }\textbf {\bibinfo {volume} {112}},\ \bibinfo
  {pages} {077204} (\bibinfo {year} {2014})}\BibitemShut {NoStop}%
\bibitem [{\citenamefont {Shankar}\ \emph {et~al.}(2014)\citenamefont
  {Shankar}, \citenamefont {Kim},\ and\ \citenamefont {Kee}}]{Shankar2014}%
  \BibitemOpen
  \bibfield  {author} {\bibinfo {author} {\bibfnamefont {V.~V.}\ \bibnamefont
  {Shankar}}, \bibinfo {author} {\bibfnamefont {H.-S.}\ \bibnamefont {Kim}}, \
  and\ \bibinfo {author} {\bibfnamefont {H.-Y.}\ \bibnamefont {Kee}},\ }\href
  {http://arxiv.org/abs/1411.6623} {\bibfield  {journal} {\bibinfo  {journal}
  {arXiv:1411.6623}\ } (\bibinfo {year} {2014})}\BibitemShut {NoStop}%
\bibitem [{\citenamefont {Foyevtsova}\ \emph {et~al.}(2013)\citenamefont
  {Foyevtsova}, \citenamefont {Jeschke}, \citenamefont {Mazin}, \citenamefont
  {Khomskii},\ and\ \citenamefont {Valent\'{i}}}]{Katerina2013}%
  \BibitemOpen
  \bibfield  {author} {\bibinfo {author} {\bibfnamefont {K.}~\bibnamefont
  {Foyevtsova}}, \bibinfo {author} {\bibfnamefont {H.~O.}\ \bibnamefont
  {Jeschke}}, \bibinfo {author} {\bibfnamefont {I.~I.}\ \bibnamefont {Mazin}},
  \bibinfo {author} {\bibfnamefont {D.~I.}\ \bibnamefont {Khomskii}}, \ and\
  \bibinfo {author} {\bibfnamefont {R.}~\bibnamefont {Valent\'{i}}},\ }\href
  {\doibase 10.1103/PhysRevB.88.035107} {\bibfield  {journal} {\bibinfo
  {journal} {Phys. Rev. B}\ }\textbf {\bibinfo {volume} {88}},\ \bibinfo
  {pages} {035107} (\bibinfo {year} {2013})}\BibitemShut {NoStop}%
\bibitem [{\citenamefont {Yamaji}\ \emph {et~al.}(2014)\citenamefont {Yamaji},
  \citenamefont {Nomura}, \citenamefont {Kurita}, \citenamefont {Arita},\ and\
  \citenamefont {Imada}}]{Yamaji2014}%
  \BibitemOpen
  \bibfield  {author} {\bibinfo {author} {\bibfnamefont {Y.}~\bibnamefont
  {Yamaji}}, \bibinfo {author} {\bibfnamefont {Y.}~\bibnamefont {Nomura}},
  \bibinfo {author} {\bibfnamefont {M.}~\bibnamefont {Kurita}}, \bibinfo
  {author} {\bibfnamefont {R.}~\bibnamefont {Arita}}, \ and\ \bibinfo {author}
  {\bibfnamefont {M.}~\bibnamefont {Imada}},\ }\href {\doibase
  10.1103/PhysRevLett.113.107201} {\bibfield  {journal} {\bibinfo  {journal}
  {Phys. Rev. Lett.}\ }\textbf {\bibinfo {volume} {113}},\ \bibinfo {pages}
  {107201} (\bibinfo {year} {2014})}\BibitemShut {NoStop}%
\bibitem [{\citenamefont {Kim}\ \emph {et~al.}(2015)\citenamefont {Kim},
  \citenamefont {Lee},\ and\ \citenamefont {Kim}}]{Kim2015}%
  \BibitemOpen
  \bibfield  {author} {\bibinfo {author} {\bibfnamefont {H.-S.}\ \bibnamefont
  {Kim}}, \bibinfo {author} {\bibfnamefont {E.~K.-H.}\ \bibnamefont {Lee}}, \
  and\ \bibinfo {author} {\bibfnamefont {Y.~B.}\ \bibnamefont {Kim}},\ }\href
  {\doibase 10.1209/0295-5075/112/67004} {\bibfield  {journal} {\bibinfo
  {journal} {Europhys. Lett.}\ }\textbf {\bibinfo {volume} {112}},\ \bibinfo
  {pages} {67004} (\bibinfo {year} {2015})}\BibitemShut {NoStop}%
\bibitem [{\citenamefont {Winter}\ \emph {et~al.}(2016)\citenamefont {Winter},
  \citenamefont {Li}, \citenamefont {Jeschke},\ and\ \citenamefont
  {Valent\'{\i}}}]{Winter2016}%
  \BibitemOpen
  \bibfield  {author} {\bibinfo {author} {\bibfnamefont {S.~M.}\ \bibnamefont
  {Winter}}, \bibinfo {author} {\bibfnamefont {Y.}~\bibnamefont {Li}}, \bibinfo
  {author} {\bibfnamefont {H.~O.}\ \bibnamefont {Jeschke}}, \ and\ \bibinfo
  {author} {\bibfnamefont {R.}~\bibnamefont {Valent\'{\i}}},\ }\href {\doibase
  10.1103/PhysRevB.93.214431} {\bibfield  {journal} {\bibinfo  {journal} {Phys.
  Rev. B}\ }\textbf {\bibinfo {volume} {93}},\ \bibinfo {pages} {214431}
  (\bibinfo {year} {2016})}\BibitemShut {NoStop}%
\bibitem [{\citenamefont {Kim}\ and\ \citenamefont {Kee}(2016)}]{Kee2016}%
  \BibitemOpen
  \bibfield  {author} {\bibinfo {author} {\bibfnamefont {H.-S.}\ \bibnamefont
  {Kim}}\ and\ \bibinfo {author} {\bibfnamefont {H.-Y.}\ \bibnamefont {Kee}},\
  }\href {\doibase 10.1103/PhysRevB.93.155143} {\bibfield  {journal} {\bibinfo
  {journal} {Phys. Rev. B}\ }\textbf {\bibinfo {volume} {93}},\ \bibinfo
  {pages} {155143} (\bibinfo {year} {2016})}\BibitemShut {NoStop}%
\bibitem [{\citenamefont {Kim}\ \emph {et~al.}(2016)\citenamefont {Kim},
  \citenamefont {Kim},\ and\ \citenamefont {Kee}}]{Kim2016}%
  \BibitemOpen
  \bibfield  {author} {\bibinfo {author} {\bibfnamefont {H.-S.}\ \bibnamefont
  {Kim}}, \bibinfo {author} {\bibfnamefont {Y.~B.}\ \bibnamefont {Kim}}, \ and\
  \bibinfo {author} {\bibfnamefont {H.-Y.}\ \bibnamefont {Kee}},\ }\href
  {https://arxiv.org/abs/1608.04741} {\bibfield  {journal} {\bibinfo  {journal}
  {arXiv:1608.04741}\ } (\bibinfo {year} {2016})}\BibitemShut {NoStop}%
\bibitem [{\citenamefont {Schaffer}\ \emph {et~al.}(2012)\citenamefont
  {Schaffer}, \citenamefont {Bhattacharjee},\ and\ \citenamefont
  {Kim}}]{Schaffer2012}%
  \BibitemOpen
  \bibfield  {author} {\bibinfo {author} {\bibfnamefont {R.}~\bibnamefont
  {Schaffer}}, \bibinfo {author} {\bibfnamefont {S.}~\bibnamefont
  {Bhattacharjee}}, \ and\ \bibinfo {author} {\bibfnamefont {Y.~B.}\
  \bibnamefont {Kim}},\ }\href {\doibase 10.1103/PhysRevB.86.224417} {\bibfield
   {journal} {\bibinfo  {journal} {Phys. Rev. B}\ }\textbf {\bibinfo {volume}
  {86}},\ \bibinfo {pages} {224417} (\bibinfo {year} {2012})}\BibitemShut
  {NoStop}%
\bibitem [{\citenamefont {Lee}\ \emph {et~al.}(2014)\citenamefont {Lee},
  \citenamefont {Schaffer}, \citenamefont {Bhattacharjee},\ and\ \citenamefont
  {Kim}}]{Lee2014}%
  \BibitemOpen
  \bibfield  {author} {\bibinfo {author} {\bibfnamefont {E.~K.-H.}\
  \bibnamefont {Lee}}, \bibinfo {author} {\bibfnamefont {R.}~\bibnamefont
  {Schaffer}}, \bibinfo {author} {\bibfnamefont {S.}~\bibnamefont
  {Bhattacharjee}}, \ and\ \bibinfo {author} {\bibfnamefont {Y.~B.}\
  \bibnamefont {Kim}},\ }\href {\doibase 10.1103/PhysRevB.89.045117} {\bibfield
   {journal} {\bibinfo  {journal} {Phys. Rev. B}\ }\textbf {\bibinfo {volume}
  {89}},\ \bibinfo {pages} {045117} (\bibinfo {year} {2014})}\BibitemShut
  {NoStop}%
\bibitem [{\citenamefont {Nishimoto}\ \emph {et~al.}(2016)\citenamefont
  {Nishimoto}, \citenamefont {Katukuri}, \citenamefont {Yushankhai},
  \citenamefont {Stoll}, \citenamefont {Roessler}, \citenamefont {Hozoi},
  \citenamefont {Rousochatzakis},\ and\ \citenamefont {van~den
  Brink}}]{Satoshi2016}%
  \BibitemOpen
  \bibfield  {author} {\bibinfo {author} {\bibfnamefont {S.}~\bibnamefont
  {Nishimoto}}, \bibinfo {author} {\bibfnamefont {V.~M.}\ \bibnamefont
  {Katukuri}}, \bibinfo {author} {\bibfnamefont {V.}~\bibnamefont
  {Yushankhai}}, \bibinfo {author} {\bibfnamefont {H.}~\bibnamefont {Stoll}},
  \bibinfo {author} {\bibfnamefont {U.~K.}\ \bibnamefont {Roessler}}, \bibinfo
  {author} {\bibfnamefont {L.}~\bibnamefont {Hozoi}}, \bibinfo {author}
  {\bibfnamefont {I.}~\bibnamefont {Rousochatzakis}}, \ and\ \bibinfo {author}
  {\bibfnamefont {J.}~\bibnamefont {van~den Brink}},\ }\href {\doibase
  10.1038/ncomms10273} {\bibfield  {journal} {\bibinfo  {journal} {Nat Commun}\
  }\textbf {\bibinfo {volume} {7}} (\bibinfo {year} {2016}),\
  10.1038/ncomms10273}\BibitemShut {NoStop}%
\bibitem [{\citenamefont {Katukuri}\ \emph {et~al.}(2015)\citenamefont
  {Katukuri}, \citenamefont {Satoshi}, \citenamefont {Rousochatzakis},
  \citenamefont {Stoll}, \citenamefont {van~den Brink},\ and\ \citenamefont
  {Hozoi}}]{Katukuri2015}%
  \BibitemOpen
  \bibfield  {author} {\bibinfo {author} {\bibfnamefont {V.~M.}\ \bibnamefont
  {Katukuri}}, \bibinfo {author} {\bibfnamefont {N.}~\bibnamefont {Satoshi}},
  \bibinfo {author} {\bibfnamefont {I.}~\bibnamefont {Rousochatzakis}},
  \bibinfo {author} {\bibfnamefont {H.}~\bibnamefont {Stoll}}, \bibinfo
  {author} {\bibfnamefont {J.}~\bibnamefont {van~den Brink}}, \ and\ \bibinfo
  {author} {\bibfnamefont {L.}~\bibnamefont {Hozoi}},\ }\href {\doibase
  10.1038/srep14718} {\bibfield  {journal} {\bibinfo  {journal} {Sci Rep}\ }
  (\bibinfo {year} {2015}),\ 10.1038/srep14718}\BibitemShut {NoStop}%
\bibitem [{\citenamefont {Rousochatzakis}\ \emph {et~al.}(2015)\citenamefont
  {Rousochatzakis}, \citenamefont {Reuther}, \citenamefont {Thomale},
  \citenamefont {Rachel},\ and\ \citenamefont {Perkins}}]{Ioannis2015}%
  \BibitemOpen
  \bibfield  {author} {\bibinfo {author} {\bibfnamefont {I.}~\bibnamefont
  {Rousochatzakis}}, \bibinfo {author} {\bibfnamefont {J.}~\bibnamefont
  {Reuther}}, \bibinfo {author} {\bibfnamefont {R.}~\bibnamefont {Thomale}},
  \bibinfo {author} {\bibfnamefont {S.}~\bibnamefont {Rachel}}, \ and\ \bibinfo
  {author} {\bibfnamefont {N.~B.}\ \bibnamefont {Perkins}},\ }\href {\doibase
  10.1103/PhysRevX.5.041035} {\bibfield  {journal} {\bibinfo  {journal} {Phys.
  Rev. X}\ }\textbf {\bibinfo {volume} {5}},\ \bibinfo {pages} {041035}
  (\bibinfo {year} {2015})}\BibitemShut {NoStop}%
\bibitem [{\citenamefont {B\"uchner}(2016)}]{Buchner2016}%
  \BibitemOpen
  \bibfield  {author} {\bibinfo {author} {\bibfnamefont {B.}~\bibnamefont
  {B\"uchner}},\ }\href@noop {} {\bibfield  {journal} {\bibinfo  {journal}
  {Talk at International FSB Conference on Frustration and Topology, Kloster
  Nimbschen, Germany}\ } (\bibinfo {year} {2016})}\BibitemShut {NoStop}%
\bibitem [{\citenamefont {Takagi}(2016)}]{Takagi2016}%
  \BibitemOpen
  \bibfield  {author} {\bibinfo {author} {\bibfnamefont {H.}~\bibnamefont
  {Takagi}},\ }\href@noop {} {\bibfield  {journal} {\bibinfo  {journal} {Talk
  at International Conference on Highly Frustrated Magnetism, Taipei, Taiwan}\
  } (\bibinfo {year} {2016})}\BibitemShut {NoStop}%
\bibitem [{\citenamefont {Yadav}\ \emph {et~al.}()\citenamefont {Yadav},
  \citenamefont {Hozoi},\ and\ \citenamefont {Tsirlin}}]{Liviu2016}%
  \BibitemOpen
  \bibfield  {author} {\bibinfo {author} {\bibfnamefont {R.}~\bibnamefont
  {Yadav}}, \bibinfo {author} {\bibfnamefont {L.}~\bibnamefont {Hozoi}}, \ and\
  \bibinfo {author} {\bibfnamefont {A.}~\bibnamefont {Tsirlin}},\ }\href@noop
  {} {\ }\bibinfo {note} {Private communication}\BibitemShut {NoStop}%
\bibitem [{\citenamefont {Lee}\ \emph {et~al.}(2016)\citenamefont {Lee},
  \citenamefont {Rau},\ and\ \citenamefont {Kim}}]{Lee2016}%
  \BibitemOpen
  \bibfield  {author} {\bibinfo {author} {\bibfnamefont {E.~K.-H.}\
  \bibnamefont {Lee}}, \bibinfo {author} {\bibfnamefont {J.~G.}\ \bibnamefont
  {Rau}}, \ and\ \bibinfo {author} {\bibfnamefont {Y.~B.}\ \bibnamefont
  {Kim}},\ }\href {\doibase 10.1103/PhysRevB.93.184420} {\bibfield  {journal}
  {\bibinfo  {journal} {Phys. Rev. B}\ }\textbf {\bibinfo {volume} {93}},\
  \bibinfo {pages} {184420} (\bibinfo {year} {2016})}\BibitemShut {NoStop}%
\bibitem [{SM()}]{SM}%
  \BibitemOpen
  \href@noop {} {}\bibinfo {note} {See Supplemental material 
  which includes \cite{Klein1974}, for auxiliary information on: i) the form of
  the interaction matrix $\bs{\Lambda}_{\vec{k}}$, ii) some special members of
  the classical ground state manifold, iii) the symmetries of the quantum
  model, and iv) technical details and additional expressions from
  RSPT.}\BibitemShut {Stop}%
\bibitem [{Note1()}]{Note1}%
  \BibitemOpen
  \bibinfo {note} {Note that here, in contrast to the Kitaev model, the fluxes
  are not conserved quantities, and so they cannot be fixed independently from
  each other.}\BibitemShut {Stop}%
\bibitem [{\citenamefont {Batista}\ and\ \citenamefont
  {Nussinov}(2005)}]{Batista05}%
  \BibitemOpen
  \bibfield  {author} {\bibinfo {author} {\bibfnamefont {C.~D.}\ \bibnamefont
  {Batista}}\ and\ \bibinfo {author} {\bibfnamefont {Z.}~\bibnamefont
  {Nussinov}},\ }\href {\doibase 10.1103/PhysRevB.72.045137} {\bibfield
  {journal} {\bibinfo  {journal} {Phys. Rev. B}\ }\textbf {\bibinfo {volume}
  {72}},\ \bibinfo {pages} {045137} (\bibinfo {year} {2005})}\BibitemShut
  {NoStop}%
\bibitem [{\citenamefont {Nussinov}\ \emph {et~al.}(2006)\citenamefont
  {Nussinov}, \citenamefont {Batista},\ and\ \citenamefont
  {Fradkin}}]{Nussinov2006}%
  \BibitemOpen
  \bibfield  {author} {\bibinfo {author} {\bibfnamefont {Z.}~\bibnamefont
  {Nussinov}}, \bibinfo {author} {\bibfnamefont {C.~D.}\ \bibnamefont
  {Batista}}, \ and\ \bibinfo {author} {\bibfnamefont {E.}~\bibnamefont
  {Fradkin}},\ }\href {\doibase 10.1142/S0217979206036326} {\bibfield
  {journal} {\bibinfo  {journal} {Int. J. Mod. Phys. B}\ }\textbf {\bibinfo
  {volume} {20}},\ \bibinfo {pages} {5239} (\bibinfo {year}
  {2006})}\BibitemShut {NoStop}%
\bibitem [{\citenamefont {Nussinov}\ and\ \citenamefont {van~den
  Brink}(2015)}]{Nussinov2015}%
  \BibitemOpen
  \bibfield  {author} {\bibinfo {author} {\bibfnamefont {Z.}~\bibnamefont
  {Nussinov}}\ and\ \bibinfo {author} {\bibfnamefont {J.}~\bibnamefont {van~den
  Brink}},\ }\href {\doibase 10.1103/RevModPhys.87.1} {\bibfield  {journal}
  {\bibinfo  {journal} {Rev. Mod. Phys.}\ }\textbf {\bibinfo {volume} {87}},\
  \bibinfo {pages} {1} (\bibinfo {year} {2015})}\BibitemShut {NoStop}%
\bibitem [{\citenamefont {Lindg\aa{}rd}(1988)}]{Lindgard1988}%
  \BibitemOpen
  \bibfield  {author} {\bibinfo {author} {\bibfnamefont {P.-A.}\ \bibnamefont
  {Lindg\aa{}rd}},\ }\href {\doibase 10.1103/PhysRevLett.61.629} {\bibfield
  {journal} {\bibinfo  {journal} {Phys. Rev. Lett.}\ }\textbf {\bibinfo
  {volume} {61}},\ \bibinfo {pages} {629} (\bibinfo {year} {1988})}\BibitemShut
  {NoStop}%
\bibitem [{\citenamefont {Long}(1989)}]{Long1989}%
  \BibitemOpen
  \bibfield  {author} {\bibinfo {author} {\bibfnamefont {M.~W.}\ \bibnamefont
  {Long}},\ }\href {http://stacks.iop.org/0953-8984/1/i=17/a=008} {\bibfield
  {journal} {\bibinfo  {journal} {Journal of Physics: Condensed Matter}\
  }\textbf {\bibinfo {volume} {1}},\ \bibinfo {pages} {2857} (\bibinfo {year}
  {1989})}\BibitemShut {NoStop}%
\bibitem [{\citenamefont {Heinil\"a}\ and\ \citenamefont
  {Oja}(1993)}]{Heinila1993}%
  \BibitemOpen
  \bibfield  {author} {\bibinfo {author} {\bibfnamefont {M.~T.}\ \bibnamefont
  {Heinil\"a}}\ and\ \bibinfo {author} {\bibfnamefont {A.~S.}\ \bibnamefont
  {Oja}},\ }\href {\doibase 10.1103/PhysRevB.48.7227} {\bibfield  {journal}
  {\bibinfo  {journal} {Phys. Rev. B}\ }\textbf {\bibinfo {volume} {48}},\
  \bibinfo {pages} {7227} (\bibinfo {year} {1993})}\BibitemShut {NoStop}%
\bibitem [{\citenamefont {Chernyshev}\ and\ \citenamefont
  {Zhitomirsky}(2014)}]{Mike2014}%
  \BibitemOpen
  \bibfield  {author} {\bibinfo {author} {\bibfnamefont {A.~L.}\ \bibnamefont
  {Chernyshev}}\ and\ \bibinfo {author} {\bibfnamefont {M.~E.}\ \bibnamefont
  {Zhitomirsky}},\ }\href {\doibase 10.1103/PhysRevLett.113.237202} {\bibfield
  {journal} {\bibinfo  {journal} {Phys. Rev. Lett.}\ }\textbf {\bibinfo
  {volume} {113}},\ \bibinfo {pages} {237202} (\bibinfo {year}
  {2014})}\BibitemShut {NoStop}%
\bibitem [{\citenamefont {Wannier}(1950)}]{Wannier1950}%
  \BibitemOpen
  \bibfield  {author} {\bibinfo {author} {\bibfnamefont {G.~H.}\ \bibnamefont
  {Wannier}},\ }\href {\doibase 10.1103/PhysRev.79.357} {\bibfield  {journal}
  {\bibinfo  {journal} {Phys. Rev.}\ }\textbf {\bibinfo {volume} {79}},\
  \bibinfo {pages} {357} (\bibinfo {year} {1950})}\BibitemShut {NoStop}%
\bibitem [{\citenamefont {Houtappel}(1950)}]{Houtappel1950}%
  \BibitemOpen
  \bibfield  {author} {\bibinfo {author} {\bibfnamefont {R.}~\bibnamefont
  {Houtappel}},\ }\href {\doibase
  http://dx.doi.org/10.1016/0031-8914(50)90130-3} {\bibfield  {journal}
  {\bibinfo  {journal} {Physica}\ }\textbf {\bibinfo {volume} {16}},\ \bibinfo
  {pages} {425 } (\bibinfo {year} {1950})}\BibitemShut {NoStop}%
\bibitem [{\citenamefont {Rousochatzakis}\ and\ \citenamefont
  {Perkins}()}]{IoannisLong}%
  \BibitemOpen
  \bibfield  {author} {\bibinfo {author} {\bibfnamefont {I.}~\bibnamefont
  {Rousochatzakis}}\ and\ \bibinfo {author} {\bibfnamefont {N.~B.}\
  \bibnamefont {Perkins}},\ }\href@noop {} {\ }\bibinfo {note} {In
  preparation}\BibitemShut {NoStop}%
\bibitem [{\citenamefont {Klein}(1974)}]{Klein1974}%
  \BibitemOpen
  \bibfield  {author} {\bibinfo {author} {\bibfnamefont {D.~J.}\ \bibnamefont
  {Klein}},\ }\href {\doibase http://dx.doi.org/10.1063/1.1682018} {\bibfield
  {journal} {\bibinfo  {journal} {The Journal of Chemical Physics}\ }\textbf
  {\bibinfo {volume} {61}},\ \bibinfo {pages} {786} (\bibinfo {year}
  {1974})}\BibitemShut {NoStop}%
\end{thebibliography}

%

\clearpage
\appendix
\pagenumbering{roman}
\begin{widetext}

\section{Supplemental material}
In this Supplementing material:
i) we provide the form of the classical energy in momentum space (Sec.~\ref{Sec:LT}) and comment on the lower energy bound and the conditions satisfied by the ground states;  
ii) we discuss three special members of the classical ground state manifold (Sec.~\ref{Sec:Special}); 
iii) we discuss three important symmetries of the quantum model that constraint the form of the effective interactions between the $\eta$ variables (Sec.~\ref{Sec:Sym}); 
iv) we give some technical details and derivations for the real space perturbation theory (Sec.~\ref{Sec:RSPT}).

\subsection{Classical energy in momentum space for the 2D honeycomb case}\label{Sec:LT}
{\cbl{\it Interaction matrix for the 2D honeycomb case--}} 
Figure~\ref{fig:Model} shows the honeycomb lattice with our convention for the primitive translations $\vec{t}_1$ and $\vec{t}_2$.  
Each site $i$ is represented as $(\vec{r},\nu)$, where $\vec{r}$ labels the position of the unit cell and $\nu\!=\!1$-$2$ is the sublattice index, see Fig.~\ref{fig:Model}. The total energy in momentum space is given by
\be\label{eq:Hq}
\mc{H}/N=\frac{1}{2}\sum_{\vec{k}} \left(\vec{S}_{-\vec{k},1}^T,\vec{S}_{-\vec{k},2}^T\right) \cdot 
\bs{\Lambda}_{\vec{k}} \cdot 
\left(\begin{array}{c}
\vec{S}_{\vec{k},1} \\
\vec{S}_{\vec{k},2}
\end{array}\right),
\ee
where $\vec{S}_{\vec{r},\nu}\!=\!\sum_{\vec{k}} e^{i \vec{k}\cdot\vec{r}} \vec{S}_{\vec{k},\nu}$, $\vec{S}_{\vec{k},\nu}\!=\!(S_{\vec{k},\nu}^x,S_{\vec{k},\nu}^y,S_{\vec{k},\nu}^z)^T$,  
and the 6$\times$6 interaction matrix $\bs{\Lambda}_\vec{k}$ is given by:
\be
\bs{\Lambda}_{\vec{k}} = \left(
\begin{array}{cc}
\vec{0} & \vec{B}_{\vec{k}} \\
\vec{B}^\ast_{\vec{k}} & \vec{0}
\end{array}
\right),~~
\vec{B}_{\vec{k}} \!= \!\frac{1}{2}\left(\!\!
\begin{array}{ccc}
0 \!&\!\Gamma \!&\! \Gamma e^{-i\vec{k}\cdot\vec{t}_3}\\
\Gamma \!&\! 0 \!&\! \Gamma e^{i\vec{k}\cdot\vec{t}_2}\\
\Gamma e^{-i\vec{k}\cdot\vec{t}_3} \!&\! \Gamma e^{i\vec{k}\cdot\vec{t}_2} \!&\! 0
\end{array}\!\!
\right),
\ee
\normalsize
where $\vec{t}_3\!=\!\vec{t}_1-\vec{t}_2$.
Due to the structure of $\bs{\Lambda}$, its eigenvectors satisfy the relation $\vec{V}_{-\vec{k},\alpha}\!=\!\vec{V}^\ast_{\vec{k},\alpha}$, where $\alpha\!=\!1$-$6$. 

{\cbl {\it Lower energy bound --}} 
One can show that the minimum eigenvalue $\lambda_{\text{min}}$ of $\bs{\Lambda}$ provides a lower bound for the energy per site $E/N$, as follows. We first expand the spin configuration into eigenmodes of $\bs{\Lambda}$:
\be
\left(\begin{array}{c}
\vec{S}_{\vec{r},1} \\
\vec{S}_{\vec{r},2}
\end{array}\right) = \sum_{\vec{k},\alpha} c_{\vec{k},\alpha} \vec{V}_{\vec{k},\alpha}.
\ee 
Then Eq.~(\ref{eq:Hq}) gives for the total energy per site:
\be
\frac{E}{N}=\frac{1}{2} \sum_{\vec{k},\alpha} \lambda_{\vec{k},\alpha} |c_{\vec{k},\alpha}|^2 \ge 
\frac{1}{2} \lambda_{\text{min}} \sum_{\vec{k},\alpha}  |c_{\vec{k},\alpha}|^2 ~.
\ee
The last term is fixed by the soft spin-length constraint 
\be
\sum_{\vec{r},\nu}\! \vec{S}_{\vec{r},\nu}^2 \!=\! N S^2 \!\Rightarrow\!
\sum_{\vec{k},\nu}\! \vec{S}_{\vec{k},\nu}\!\cdot\!\vec{S}_{-\vec{k},\nu} \!=\! \sum_{\vec{k},\alpha}\! |c_{\vec{k},\alpha}|^2  \!=\! 2 S^2,
\ee
which then leads to the lower bound of the energy per site:
\be
E/N \ge\lambda_{\text{min}} S^2~.
\ee

\begin{figure}[!t] 
\includegraphics[width=0.29\textwidth,angle=0,clip=true,trim=0 0 0 0]{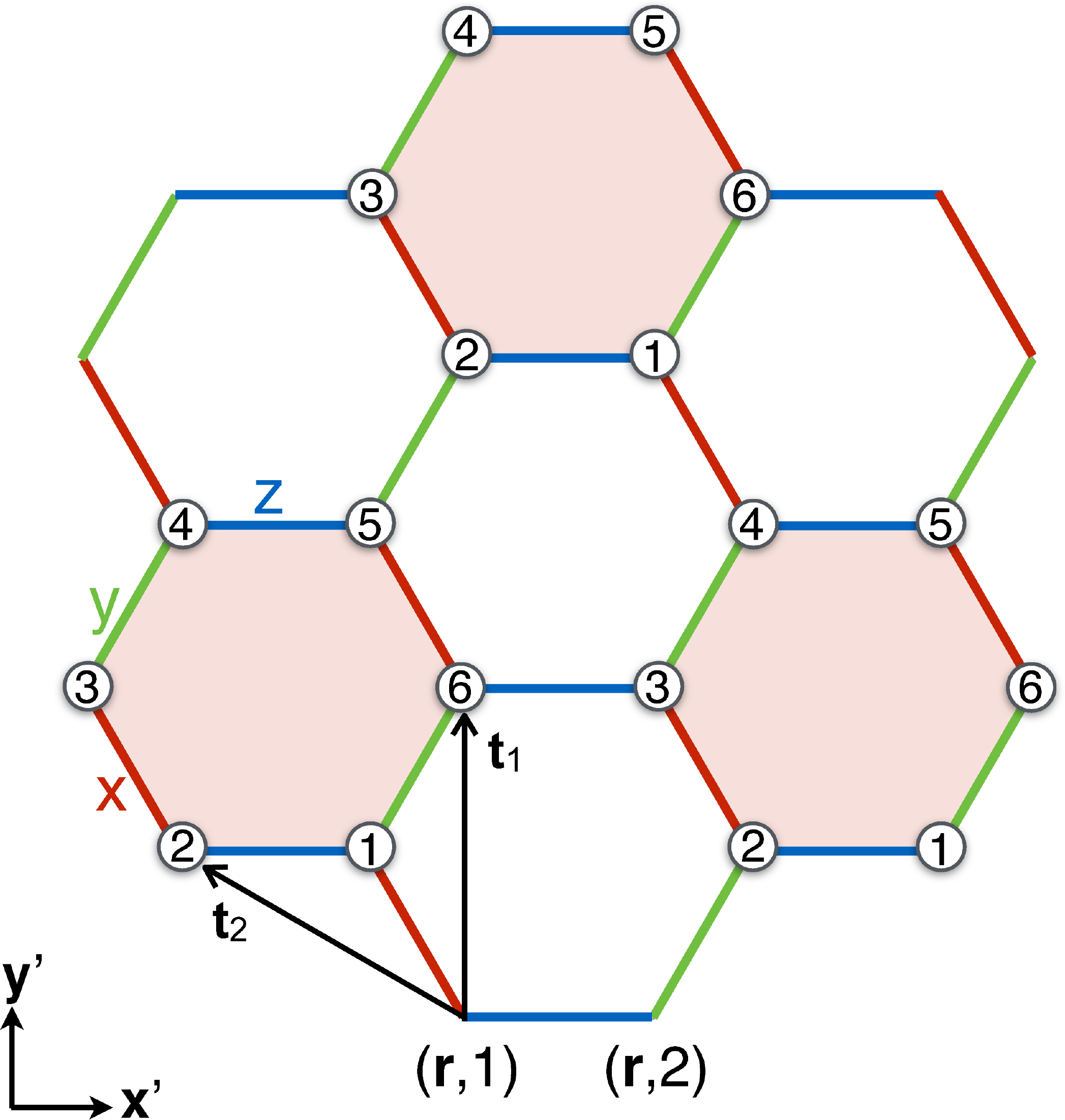}
\caption{2D honeycomb lattice, with three types of NN bonds, labeled by $\alpha\!=\!x$, $y$ or $z$. The axes $\vec{x'}$ and $\vec{y}'$ define the plane of the lattice,  $\vec{t}_1\!=\!a \vec{y}'$ and $\vec{t}_2\!=\!a(-\frac{\sqrt{3}}{2}\vec{x}'+\frac{1}{2}\vec{y}')$ are primitive translations, and $a$ is the lattice constant. The two sites of the unit cell are denoted by $(\vec{r},1)$ and $(\vec{r},2)$. The labels $1$-$6$ inside the shaded hexagons refer to the six-sublattice symmetry $\mc{R}_a$, see text.}\label{fig:Model}
\end{figure}

{\cbl{\it Conditions satisfied by the classical ground states of the $\Gamma$ model --}} 
One can use a rigorous argument based on the expansion of the energy into eigenmodes of $\bs{\Lambda}_{\vec{k}}$, to show that the ground states described in the main text exhaust all possibilities. 
We begin by noticing that since all ground states must saturate the lower energy bound $\lambda_{1}S^2$, it follows that they should be described as linear superpositions of eigenmodes of the lowest band of the interaction matrix only. 
\be\label{eq:V1}
\left(\begin{array}{c}
\vec{S}_{\vec{r},1}\\
\vec{S}_{\vec{r},2}
\end{array}\right)  = 
\sum_{\vec{k}} c_{\vec{k}} e^{i\vec{k}\cdot\vec{r}} 
\vec{V}_1(\vec{k})~,
\ee
where $\vec{V}_1\!=\!(\vec{u},\vec{w})$ is the eigenvector corresponding to $\lambda_1$. For positive $\Gamma$, this takes the form:
\bea
\vec{u}=( e^{-\frac{i}{2} (\sqrt{3}k_x+k_y)}, e^{-\frac{i}{2} (\sqrt{3}k_x-k_y)},e^{-i \frac{\sqrt{3}}{2}k_x} ),~~~~
\vec{w}=( -e^{-\frac{i}{2} (\sqrt{3}k_x-k_y)}, -e^{-\frac{i}{2} (\sqrt{3}k_x+k_y)},-1)~.
\eea
This form shows that, irrespectively of the coefficients $c_{\vec{k}}$ in (\ref{eq:V1}), we have (again, for positive $\Gamma$): 
\bea
&&x\text{-bonds}: ~~S_{\vec{r},2}^y\!=\!-S_{\vec{r}+\vec{t}_3,1}^z,~~~
S_{\vec{r},2}^z\!=\!-S_{\vec{r}+\vec{t}_3,1}^y, \label{eq:x}\\
&&y\text{-bonds}: ~~S_{\vec{r},2}^x\!=\!-S_{\vec{r}-\vec{t}_2,1}^z,~~~
S_{\vec{r},2}^z\!=\!-S_{\vec{r}-\vec{t}_2,1}^x\label{eq:y},\\
&&z\text{-bonds}: ~~S_{\vec{r},2}^x\!=\!-S_{\vec{r},1}^y,~~~
S_{\vec{r},2}^y\!=\!-S_{\vec{r},1}^x, \label{eq:z}
\eea
which are precisely the conditions satisfied by the components of the spins on a $x$-type (\ref{eq:x}),  $y$-type (\ref{eq:y}), or $z$-type (\ref{eq:z}) of bonds, as described in the main text. 
So the states described in the main text exhaust all possible ground states of the $\Gamma$-model.

\subsection{Special members of the ground state manifold}\label{Sec:Special}
Here we briefly discuss three special members of the ground state manifold for $\Gamma\!>\!0$. 
The classical ground states of the $\Gamma\!<\!0$ case can be obtained by time reversal operation in every second lattice site. 
For simplicity, we consider the 2D honeycomb case, and analogous states exist for the 3D cases as well.

One special family inside the ground state manifold are the $2^{N/2}$ states corresponding to $a\!=\!b\!=\!c\!=\!\frac{S}{\sqrt{3}}$. Two members of this family are the N\'eel state along $[111]$ and the zigzag states along the $\langle\bar{1}11\rangle$ axes.   

A second special family of ground states are the $2^{N/6}$ states corresponding to $a\!=\!S$ and $b\!=c\!=\!0$, which is shown in Fig.~\ref{fig:TwoSpecialStates}~(a). Here, each spin points along one of the cubic axes, and the energy comes solely from the interactions within the A-type (shaded) hexagons. The $2^{N/6}$ states arise by applying the time-reversal operation to the six spins of any of the A-type hexagons.

A third special family consists of the $2^{N/3}$ ground states corresponding to $a\!=\!b\!=\!\frac{S}{\sqrt{2}}$ and $c\!=\!0$. A special member of this family is shown in Fig.~\ref{fig:TwoSpecialStates}~(b). Here, the spins form FM dimers pointing along one the face diagonals,  $[1\bar{1}0]$, $[\bar{1}01]$ or $[01\bar{1}]$, depending on whether the dimers sit on a $z$, $y$ or $x$ bond, respectively. In this state, each intra-dimer coupling contributes an energy of $-|\Gamma| S^2$, while each inter-dimer coupling gives a contribution of $-|\Gamma| S^2/2$. 

\begin{figure*}[!t] 
\includegraphics[width=0.7\textwidth,angle=0,clip=true,trim=0 0 0 0]{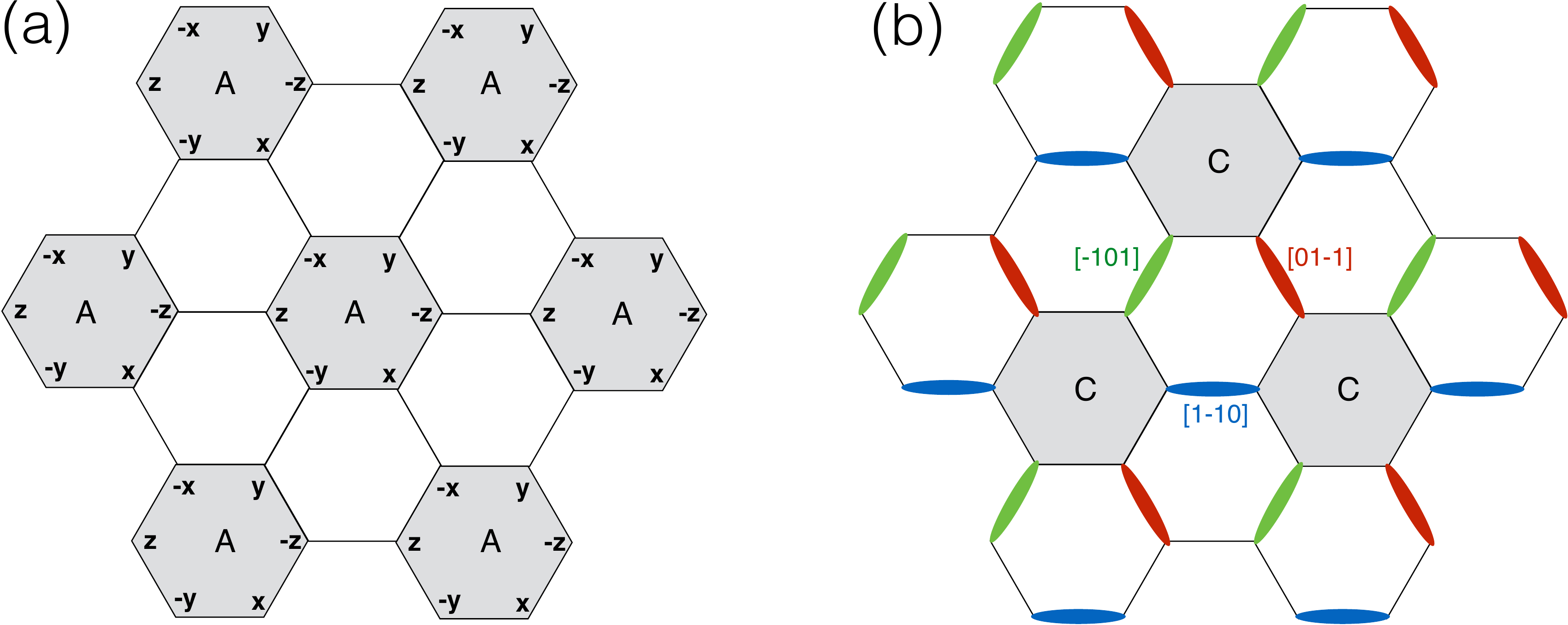}
\caption{Two special members of the ground state manifold of the $\Gamma$-model ($r\!=\!0$). (a) One of the $2^{N/6}$ ground states generated by $|a|\!=S$, $\!b\!=\!c\!=\!0$, and spins pointing along the cubic axes. (b) One of the $2^{N/3}$ ground states generated by $|a|\!=\!|b|\!=\!\frac{S}{\sqrt{2}}$, with ferromagnetic dimers (denoted by ovals) along the face diagonals $[1\bar{1}0]$ (blue), $[\bar{1}01]$ (green), and $[01\bar{1}]$ (red). The shaded hexagons show the trimerization of the lattice by these types of ground states.}\label{fig:TwoSpecialStates}
\end{figure*}

\subsection{Symmetries of the quantum model and effective interactions between {\large $\eta$} variables}\label{Sec:Sym}
The quantum $\Gamma$ model has three global, gauge-like symmetries operations $\mc{R}_a$, $\mc{R}_b$ and $\mc{R}_c$. Each one is associated with a specific six-sublattice decomposition of the honeycomb lattice. For $\mc{R}_a$, the decomposition is shown in Fig.~\ref{fig:Model} by shaded hexagons, and 
\bea\label{eq:Ra}
\mc{R}_a = \!\!
\!\!\prod_{i\in \{1, 4\}}\!\!\mathsf{C}_{2x}(i)
\!\!\prod_{i'\in \{2, 5\}}\!\!\mathsf{C}_{2y}(i')
\!\!\prod_{i''\in \{3, 6\}}\!\!\mathsf{C}_{2z}(i'')~,
\eea
where $C_{2\alpha}$ denotes a 180$^\circ$-rotation in spin space around the $\alpha$-th axis. The operations $\mc{R}_b$ and $\mc{R}_c$ look exactly the same but the labeling of the sites corresponds to the remaining two ways to choose the shaded hexagons of Fig.~\ref{fig:Model}. 

We now show that the operation $\mc{R}_a$ combined with time reversal $\mc{T}$ amounts to flipping the signs of all $\eta$ variables of the A-type. Consider the three $\eta$'s of the A-type represented by the shaded hexagons in Fig.~\ref{fig:Model}. For any such hexagons, the components of the six spins transform as follows under $\mc{R}_a \cdot \mc{T}$:
We have: 
\bea
&&
\vec{S}'_1 = ({\cred-S_1^x},S_1^y,S_1^z),~~
\vec{S}'_2 = (S_2^x,{\cred-S_2^y},S_2^z),~~
\vec{S}'_3 = (S_3^x,S_3^y,{\cred-S_3^z}), \nonumber\\
&&
\vec{S}'_4 = ({\cred-S_4^x},S_4^y,S_4^z),~~
\vec{S}'_5 = (S_5^x,{\cred-S_5^y},S_5^z),~~
\vec{S}'_6 = (S_6^x,S_6^y,{\cred-S_6^z}).
\eea
The components that change sign are precisely the ones involved in the definition of the associated ${\cred\eta}$ variable, so $\mc{R}_a\cdot\mc{T}$ amounts to flipping all $\eta$'s of the A-type. Similarly, $\mc{R}_b\cdot\mc{T}$ and $\mc{R}_c\cdot\mc{T}$ flip the signs of all $\eta$'s that belong to B- and C- type, respectively. 

Since the above operations are symmetries of the quantum Hamiltonian, it follows that the effective interactions between the $\eta$ variables, that are generated by quantum fluctuations, must respect the symmetries as well. As a result, terms that contain an odd number of $\eta$'s of the same type are excluded from the effective model. 
For example, the only bilinear terms of the type $\eta\eta'$ that are allowed are the ones where both $\eta$ and $\eta'$ belong to the same type. 
So, to leading order, different sublattices are decoupled from each other. 
The first type of processes that involve interactions between different types of $\eta$'s arise in fourth-order of perturbation theory, and are of the form $(\eta_1\eta_2)(\eta_3\eta_4)$, where $(\eta_1,\eta_2)$ belong to one type and $(\eta_3,\eta_4)$ belong to another, see main text.

\subsection{Real space perturbation theory (RSPT)}\label{Sec:RSPT}
{\cbl{\it General setting of RSPT --}} 
Consider a ground state of the $\Gamma$ model, where each spin points along a local axis $\vec{e}_i^z$. Define two perpendicular axes $\vec{e}_i^x$ and $\vec{e}_i^y$ and write
\bea
&&\vec{S}_i 
= S_i^z \vec{e}_i^z+S_i^+ \vec{e}_i^- + S_i^- \vec{e}_i^+,~~~\text{where}~~\vec{e}_i^\pm = \frac{1}{2}(\vec{e}_i^x\pm i \vec{e}_i^y).
\eea
Next we write the general form of the Hamiltonian as
\small
\bea
\mc{H} &=& \frac{1}{2}\sum_{ij} \vec{S}_i\cdot\vec{A}_{ij}\cdot \vec{S}_j 
= 
\frac{1}{2}\sum_{ij} \Big( A_{ij}^{zz} S_i^z S_j^z + A_{ij}^{z+} S_i^z S_j^- + A_{ij}^{z-} S_i^z S_j^+ + A_{ij}^{+z} S_i^- S_j^z + A_{ij}^{++} S_i^- S_j^- + A_{ij}^{+-} S_i^- S_j^+ \nonumber\\
&&+
A_{ij}^{-z} S_i^+ S_j^z + A_{ij}^{-+} S_i^+ S_j^- + A_{ij}^{--} S_i^+ S_j^+ \Big),
\eea
\normalsize
where $\bf{A}$ is a second-rank tensor, which in the present case describes the off-diagonal exchange interactions, and 
\be
A_{ij}^{zz}\!=\!\vec{e}_i^z\cdot\vec{A}_{ij}\cdot\vec{e}_j^z, ~~ A_{ij}^{z+}\!=\!\vec{e}_i^z\cdot\vec{A}_{ij}\cdot\vec{e}_j^+, ~~\text{etc}.
\ee
Next, we define the deviation operator $n_i \!=\! S-S_i^z$ and rewrite
\small
\bea
\mc{H}&=&\frac{1}{2}\sum_{ij}\Big( A_{ij}^{zz} (S-n_i) (S-n_j) 
+ A_{ij}^{z+} (S-n_i) S_j^- + A_{ij}^{z-} (S-n_i) S_j^+ + A_{ij}^{+z} S_i^- (S-n_j) + A_{ij}^{++} S_i^- S_j^- + A_{ij}^{+-} S_i^- S_j^+ \nonumber\\
&&+
A_{ij}^{-z} S_i^+ (S-n_j) + A_{ij}^{-+} S_i^+ S_j^- + A_{ij}^{--} S_i^+ S_j^+ \Big)
\eea
\normalsize
Introducing the classical energy, $E_{cl}=S^2/2\sum_{ij} A_{ij}^{zz}$, and the local field $\vec{B}_j\!=\!- S \sum_{i} \vec{e}_i^z \cdot \vec{A}_{ij} = - B_j \vec{e}_j^z$, we obtain:
\small
\bea
\mc{H}\!&=&\!E_{cl}  \!+\! \sum_{j} B_j n_j 
\!+\! \frac{1}{2}\sum_{ij} 
  \left( A_{ij}^{++}\!\!\!\!\!\underbrace{S_i^- S_j^-}_{\text{{\cbl double spin-flip}}} + A_{ij}^{+-} \!\!\!\!\!\underbrace{S_i^- S_j^+}_{\text{{\cbl spin-flip hopping}}} + h.c. \right) 
-\sum_{ij}  \left( A_{ij}^{z+} \!\!\!\!\!\underbrace{n_i S_j^-}_{\text{{\cbl single spin-flip}}}  \!+\! h.c. \right)
+(\underbrace{S \sum_{i}  A_{ij}^{z+}}_{-B_j \vec{e}_j^z \cdot \vec{e}_j^+ = 0} S_j^-  \!+\! h.c. ) .
\eea
\normalsize
In the following, we set 
\small
\be\label{eq:H0V}
\boxed{\mc{H}_0\!=\!E_{cl}\!+\!\sum_{j} B_j n_j}, ~~~
\boxed{\mc{V}\!=\!\mc{H}-\mc{H}_0\!=\!\mc{V}_{1}+\mc{V}_{2}+\mc{V}_{3}},
\ee
\normalsize
where
\small
\be
\mc{V}_1 = \frac{1}{2}\sum_{ij} \left( A_{ij}^{++} S_i^- S_j^- + h.c. \right),~
\mc{V}_2 = \frac{1}{2}\sum_{ij} \left( A_{ij}^{+-}S_i^- S_j^+ + h.c. \right),~
\mc{V}_3 = -\frac{1}{2}\sum_{ij}   \left( A_{ij}^{z+} n_i S_j^- + A_{ij}^{+z} S_i^- n_j  + h.c. \right)
\ee
\normalsize
correspond to the double spin-flip processes ($\mc{V}_1$), single spin-flip hopping ($\mc{V}_2$), and correlated, single spin-flip processes ($\mc{V}_3$). The latter are analogous to the cubic magnon terms in the standard Holstein-Primakof spin-wave expansion. 

Equation (\ref{eq:H0V}) form the basis for the RSPT. In the present problem, we have pushed RSPT up to fourth-order in $\mc{V}$ and for general spin $S$. 
The second-order terms can be obtained from the standard expression
\be\label{eq:H2}
\mc{H}_{\text{eff}}^{(2)} = \langle 0 | \mc{V} \mc{R} \mc{V} |0\rangle,
\ee
where $|0\rangle$ is the ground state of $\mc{H}_0$, $E_0$ is the corresponding energy, and $\mc{R}=\frac{1-|g\rangle\langle g|}{E_0-\mc{H}_0}$ is the resolvent. 
The third-order terms vanish while the fourth-order terms are obtained from the expression:~\cite{Klein1974}
\be\label{eq:H4}
\mc{H}_{\text{eff}}^{(4)} = \langle 0 | \mc{V} \mc{R} \mc{V} \mc{R} \mc{V} \mc{R} \mc{V} |0\rangle - \langle 0 | \mc{V} \mc{R} \mc{V} |0\rangle ~ \langle 0 | \mc{V} \mc{R}^2 \mc{V}|0\rangle
\ee

Now, since we expand around the fully polarized state (in the rotated axes system), it follows that we should always begin and end with two spin-flips, i.e. with $\mc{V}_1$. 
So, for the second-order terms (and for the second term of Eq.~(\ref{eq:H4}) above) we may replace $\mc{V}$ by $\mc{V}_1$. Since each application of $\mc{V}_1$ gives an overall factor that scales linearly with $S$, while $\mc{R}$ gives a factor of $1/S$, it follows that the second-order terms are all linear in $S$. The explicit form of these terms are given in the main text.

The first term of Eq.~(\ref{eq:H4}) gives four types of contributions:

(i) First come the ones that do not involve cubic terms, and involve up to one spin-flip per site. A typical term is given by:
\be
\langle gs | S_i^+S_k^+ \mc{R} S_i^-S_j^+\mc{R} S_i^+S_k^-\mc{R} S_i^-S_j^-|gs\rangle
\ee
where $i\ne j\ne k$. These terms are linear in $S$.

(ii) Next come the terms that involve up to two spin-flips on the same site. A typical term is:
\bea
&&\langle gs | S_i^+S_k^+ \mc{R} S_i^+S_j^+\mc{R} S_i^-S_k^-\mc{R} S_i^-S_j^-|gs\rangle
\eea 
Keeping track of the matrix elements of raising and lowering operators, we find that these terms are proportional to 
$2S-1$.

(iii) Next come the terms that involve two sites only and two spin-flips on each of these sites. A typical term is:
\bea
&&\langle gs | S_i^+S_j^+ \mc{R} S_i^+S_j^+\mc{R} S_i^-S_j^-\mc{R} S_i^-S_j^-|gs\rangle
\eea 
These terms are proportional to $(2S-1)/S$.

(iv) Finally, there are the terms that involve cubic processes $\mc{V}_3$. A typical term is:
\bea
&&\langle gs | S_i^+S_j^+ \frac{1}{\mc{R}} n_j S_k^+\frac{1}{\mc{R}} n_i S_k^-\frac{1}{\mc{R}} S_i^-S_j^-|gs\rangle
\eea
These terms do not depend on $S$.

Clearly, the terms involving more than one spin-flip per site appear only for $S>1/2$. So the functional form of the effective Hamiltonians for $S=1/2$ and $S>1/2$ differ from each other.
In the main text we have provided the expressions for $S=1/2$. 
For $S>1/2$, we get the following expressions for the connected cluster of equation (1) of the main text. First, the correction to $J_{{\cred A}}$ is:
\small
\be
\delta J_{{\cred A}}=\frac{\Gamma}{128}\left\{
(2S-\frac{23}{6} + \frac{1}{4S}) {\cred\tilde{a}}^2 
-(S-\frac{22}{3} + \frac{2}{3S}) {\cred\tilde{a}}^4 
 + (S-\frac{13}{2} + \frac{11}{12S}) {\cred\tilde{a}}^6 
- (S-\frac{13}{3}) {\cred\tilde{a}}^2 {\cgr\tilde{b}}^2{\cbl\tilde{c}}^2 
 \right\}.
\ee
\normalsize
Next, the correction to $E_{\text{ani}}$ is (disregarding constants):
\small
\be
\delta E_{\text{ani}}=\frac{-|\Gamma|}{12288}\left\{
(\frac{10}{S} + 52 - 24 S) ( {\cred\tilde{a}}^4 + {\cgr\tilde{b}}^4 + {\cbl\tilde{c}}^4)  
- 8 (\frac{6}{S} - 88 + 57 S)  {\cred\tilde{a}}^2 {\cgr\tilde{b}}^2 {\cbl\tilde{c}}^2 
+ (\frac{19}{S} - 156 + 60 S) ( {\cred\tilde{a}}^8 + {\cgr\tilde{b}}^8 + {\cbl\tilde{c}}^8) 
\right\} .
\ee
\normalsize
Finally, the four-body coupling $J_{{\cred A}{\cgr B}}$ (see definition in the main text) is:
\small
\be
J_{{\cred A}{\cgr B}} = \frac{7|\Gamma|}{384} {\cred\tilde{a}}^2{\cgr\tilde{b}}^2,
\ee
\normalsize
which is independent of $S$.

\end{widetext}

\end{document}